\documentclass[12pt]{extarticle}

\usepackage{hyperref}
\pdfoutput=1
\usepackage{graphics}
\usepackage[english] {babel}
\usepackage{amssymb,amsfonts,amsmath,mathtext,mathabx}
\usepackage{graphicx}
\usepackage{xcolor}

\usepackage{soul} 
\usepackage{subfig}
\usepackage{floatrow}

\usepackage{nicefrac}

\usepackage{float}
\floatstyle{plaintop}
\restylefloat{table}

\usepackage{natbib}
\setcitestyle{authoryear,open={(},close={)}}

\DeclareMathOperator{\Ex}{\mathbb{E}} 

\newcommand{\sd}{{\mathsf{sd}}\,}

\newcommand{\Corr}{{\mathsf{Corr}}\,}
\newcommand{\Cov}{{\mathsf{Cov}}\,}

\renewcommand{\d}{{\rm d}}

\newcommand{\etc}{etc.\ }
\newcommand{\eg}{e.g.,\ }
\newcommand{\eeg}{E.g.,\ }

\newcommand{\vs}{{\em vs.}\ }
\newcommand{\ie}{i.e.,\ }

\newcommand{\e}{\mathrm{e}}
\renewcommand{\i}{{\mathsf i}}

   \title{Additive Model Perturbations Scaled by Physical Tendencies for Use in Ensemble Prediction\footnote{
        This is the authors' final version of the paper published in 
    Tellus A: Dynamic Meteorology and Oceanography.  
    DOI: \url{https://doi.org/10.16993/tellusa.3224}
        }
        }
   \author{Michael Tsyrulnikov$^{(1)}$\footnote{
   Corresponding author, mik.tsyrulnikov@gmail.com, ORCID 0000-0002-7357-334X
   }, Elena Astakhova$^{(1)}$, and Dmitry Gayfulin$^{(1)}$    \\
   \smallskip
   \small{\em $^{(1)}$HydroMetCenter of Russia, Moscow, Russia}
}

\begin{document}

\maketitle
\floatsetup[figure]{style=plain,subcapbesideposition=top}

\begin{abstract}
Imperfections and uncertainties in forecast models are often represented in ensemble prediction systems
by stochastic perturbations of model equations.
In this article, we present a new technique to generate model perturbations.
The technique is termed Additive Model-uncertainty perturbations scaled by Physical Tendencies (AMPT). 
The generated perturbations are independent between different model variables and
scaled by the local-area-averaged modulus of physical tendency. 
The previously developed Stochastic Pattern Generator 
is used to generate space and 
time-correlated pseudo-random fields. 
AMPT attempts to address some weak points of the popular model perturbation scheme 
known as Stochastically Perturbed Parametrization Tendencies (SPPT). 
Specifically, AMPT can produce non-zero perturbations even
at grid points where the physical tendency is zero and avoids perfect correlations in the perturbation fields
in the vertical and between different variables.
Due to a non-local link from physical tendency to the local perturbation magnitude, AMPT
can generate significantly greater perturbations than SPPT  without causing instabilities.
Relationships between the bias and the spread caused by AMPT and SPPT were studied 
in an ensemble of forecasts. The   non-hydrostatic, convection-permitting forecast model COSMO
was used. 
In ensemble prediction experiments,   
AMPT perturbations led to statistically significant improvements (compared to SPPT) in probabilistic 
performance scores such as spread-skill relationship, CRPS, Brier Score, and ROC area
for near-surface temperature. AMPT had similar but weaker effects on near-surface wind speed
and mixed effects on precipitation. 
\end{abstract}



\section{Introduction}
\label{Intro}

Forecasting natural phenomena such as weather cannot be perfect.
Knowing the degree of the imperfection (the forecast uncertainty)
is always desirable and sometimes vital, for example, in decision-making.
A rough estimate of forecast uncertainty can be obtained by comparing  forecasts with verifying
observations and averaging the forecast-minus-observation statistics over time  and space.
This simple approach often results in useful estimates.
However, the `climatological' estimates can be insufficient if we predict the state of a
nonlinear chaotic system like the Earth's atmosphere, where the forecast uncertainty can vary
depending on the  weather situation
(on the local structure of the atmospheric flow) and the observational coverage.

To allow for a situation-dependent assessment of  the forecast uncertainty, 
a {\em forecast of forecast uncertainty} is needed.
In probabilistic terms, we have to predict not just the state of the system (in our case, the atmosphere)
but also the {\em  probability distribution} of the unknown true atmospheric state around the forecast.
A Monte-Carlo-based approach known as ensemble prediction, in which the probability distribution of the truth
is {\em represented} by a small number (tens to hundreds) of points 
in state space (ensemble members), has proven to be both feasible and useful
in predicting major features of forecast uncertainty, see, \eg \citet{LeutbecherPalmer,Wilks}
and references therein.

For ensemble prediction to be successful in predicting the forecast uncertainty,
ensemble members  need to be pseudo-random draws from a probability distribution that is reasonably close
to the conditional probability distribution of the truth given the data available prior  to the forecast.
The most promising approach here is, arguably, to identify all {\em sources of uncertainty}
that affect the forecast and then stochastically model each of those `input' uncertainties individually.
The forecast uncertainty is caused by uncertainties in (i) meteorological observations, (ii) data assimilation techniques,
 (iii) boundary conditions, and (iv) the forecast model itself.
In this study we are concerned with the latter source, the {\em model uncertainty}.


\subsection{Model uncertainty}
\label{sec_ME}

A numerical weather prediction model computes the forecast by time stepping.
At each time step, the model has on input, 
typically, the current model state (defined on a spatial grid),
and computes the model state at the next time step or, equivalently, computes the  {\em forecast tendency},
the difference between the next-time-step and the present model states.
The {\em model uncertainty} is, by definition, 
the  uncertainty in the forecast tendency, e.g. \citet{Orrell}.
Being accumulated and transformed during the time stepping, 
the uncertainty in the tendency contributes to 
the uncertainty in the forecast fields.

The model uncertainty is caused by the following imperfections
in the forecast model (listed in order of increasing importance).
\begin{enumerate}
\item
Atmospheric model's partial differential equations normally involve a number of  simplifications like
the neglect of variations (horizontal and vertical) in the gravitational force 
or the ideal gas law assumption.
The model may also omit some processes like chemical reactions 
or development and impact of electric charges of hydrometeors.

\item
A classical (\ie  based on laws of physics, not a neural network) meteorological forecast model
solves a set of time and space-{\em discretized} differential equations.
The discretization leads to {\em truncation} error.

\item
 {\em Subgrid-scale} processes are accounted for in atmospheric models in an approximate manner.
On the one hand, these processes cannot be reproduced by the discretized model equations because
the model grid is too coarse.
On the other hand, subgrid-scale processes do impact  grid-scale fields 
due to the nonlinearity of physical laws that govern the evolution of the atmosphere.
In forecast models, this impact  is assessed 
using simplified sub-models known as physical parametrization schemes.
Simplifications (made in these schemes for computational reasons) along with the inherent uncertainty
in the unresolved scales lead to {\em uncertainties/errors in physical parametrizations}, \eg \citet{Palmer}.

Note that some processes in the atmosphere (such as turbulence, convection, and gravity waves)
can become increasingly resolved by the model equations just by refining the
computational grid, that is, by increasing the model's spatial and temporal resolution.
Therefore, these processes can, actually, be regarded as part of truncation error.
The respective physical parametrization schemes, 
thus, attempt to reduce  truncation errors while introducing their own 
(presumably, smaller) errors/uncertainties.

\end{enumerate}

Our focus in this research is on uncertainties/errors 
in physical parametrizations of subgrid-scale processes.

\subsection{Error and uncertainty}
\label{sec_ErrUncert}

At each model time step, nonlinear interactions of
the subgrid-scale field ${\bf x}_{\rm SGS}$ 
with itself and  
with the  grid-scale field ${\bf x}_{\rm GS}$
produce a spectrum of combination wavenumbers some of which fall within the resolvable (grid-scale) range.
This yields a contribution of the subgrid-scales to the (grid-scale)
forecast tendency \citep[see, e.g.,][for a  formulation of the problem in turbulence theory]{StochTurb}. 
Having a perfect model, we could compute this contribution, 
let it be denoted $\boldsymbol\pi({\bf x}_{\rm SGS},{\bf x}_{\rm GS})$. 
Deterministic physical parametrization schemes,  having, by definition, access only to ${\bf x}_{\rm GS}$,
attempt to assess this impact, producing a {\em physical tendency}, ${\bf P}({\bf x}_{\rm GS})$.
The uncertainty in all physical parametrization schemes combined is the difference between
${\bf P}({\bf x}_{\rm GS})$ and the {\em true physical tendency} $\boldsymbol\pi({\bf x}_{\rm SGS},{\bf x}_{\rm GS})$. 

The subgrid-scale field ${\bf x}_{\rm SGS}$ is not explicitly defined in the model, 
therefore it is unknown 
and even unknowable  to the model (to the extent that ${\bf x}_{\rm SGS}$ is not determined  by ${\bf x}_{\rm GS}$).
Therefore we assume that ${\bf x}_{\rm SGS}$ is a random field  with some 
conditional probability density $p({\bf x}_{\rm SGS} \mid {\bf x}_{\rm GS})$
(we condition on ${\bf x}_{\rm GS}$ because it is available to the model).
This density induces
the  probability density $p(\boldsymbol\pi \mid {\bf x}_{\rm GS})$.
Then, the ideal (best in the mean square sense) 
{\em deterministic} physical tendency is the conditional expectation of the true tendency given the grid-scale field,
\eg \citet{Kwasniok,ShuttsPalares},
${\bf P}_{\rm ideal}^{\rm determ}({\bf x}_{\rm GS}) = \Ex(\boldsymbol\pi \mid {\bf x}_{\rm GS})$,
where $\Ex$ stands for expectation. 
It is ${\bf P}_{\rm ideal}^{\rm determ}({\bf x}_{\rm GS})$ that a deterministic physical parametrization seeks to 
approximate by  ${\bf P}({\bf x}_{\rm GS})$.
In these terms, it is meaningful to call 
\begin {equation}
\label{err}
\varepsilon_{\bf P} =  {\bf P} - {\bf P}_{\rm ideal}^{\rm determ} =
   {\bf P} - \Ex(\boldsymbol\pi \mid {\bf x}_{\rm GS})
\end {equation}
 the {\em error} in the 
deterministic physical tendency ${\bf P}$.
However, the deterministic physical tendency is incapable of  simulating the variability of $\boldsymbol\pi$
around its conditional mean value $\Ex(\boldsymbol\pi \mid {\bf x}_{\rm GS})$,
\begin {equation}
\label{uncert}
\delta_{\bf P} = \boldsymbol\pi - \Ex(\boldsymbol\pi \mid {\bf x}_{\rm GS})
\end {equation}
\citep{Palmer_simulator}.
This is the {\em inherent} (irreducible)
uncertainty in $\boldsymbol\pi({\bf x}_{\rm SGS},{\bf x}_{\rm GS})$ given ${\bf x}_{\rm GS}$. 
Then, the full uncertainty 
is the difference between the error $\varepsilon_{\bf P}$ and the irreducible uncertainty $\delta_{\bf P}$:
\begin {equation}
\label{ErrUncert}
\boldsymbol\varepsilon = 
         {\bf P}-\boldsymbol\pi \equiv [{\bf P} - \Ex(\boldsymbol\pi \mid {\bf x}_{\rm GS})] - 
                         [ \boldsymbol\pi - \Ex(\boldsymbol\pi \mid {\bf x}_{\rm GS})] =
                          \boldsymbol\varepsilon_{\bf P} - \boldsymbol\delta_{\bf P}.
\end {equation}
The irreducible uncertainty $\delta_{\bf P}$ is especially important in so-called `gray zones', where 
a process is {\em partly} resolved by the model, meaning that a length scale of ${\bf x}_{\rm SGS}$
is comparable to the grid spacing. In this case, in each grid cell, the grid-scale impact
of the subgrid-scale process
is, effectively,  a sum of a {\em small} (and random) number, $\nu$, of random contributions: 
$\boldsymbol\pi = \sum_{\ell=1}^\nu \boldsymbol\pi_\ell$.
With convection, the contributions $\boldsymbol\pi_\ell$ to the convective tendency are due to 
individual convective plumes \citep{PlantCraig}.
With boundary-layer turbulence, $\boldsymbol\pi_\ell$ are due to individual turbulent eddies \citep{KoberCraig}.
Assuming, for presentation purposes, that all $\boldsymbol\pi_\ell$ are the same in a grid cell and their
number $\nu$ follows the Poisson distribution, we readily obtain that 
$|\Ex\boldsymbol\pi| /\sd\boldsymbol\pi = \sqrt{\Ex \nu}$,
where $\sd$ denotes standard deviation.
This  expression shows that the standard deviation (which measures randomness) and the mean value 
of the grid-scale impact of subgrid-scale processes 
are comparable to each other in magnitude if $\Ex\nu \sim 1$, which is the case in a gray zone as 
discussed above.
This makes the irreducible uncertainty in $\boldsymbol\pi$ indeed substantial and implies that it
needs to be properly
accounted for in ensemble prediction.
 
It is worth remarking that the  irreducible uncertainty $\delta_{\bf P}$  
caused by the randomness of $\boldsymbol\pi$ given ${\bf x}_{\rm GS}$
is the {\em aleatory} (truly random) uncertainty,
therefore it cannot be called error.
Whereas  $\varepsilon_{\bf P}$ is a kind of systematic error (reducible conditional bias),
which is caused by imperfections in the physical parametrization schemes. This latter kind of
uncertainty due to {\em lack of knowledge} is generically called {\em epistemic}.
Both aleatory and epistemic uncertainties need to be taken into account in building an
ensemble prediction scheme.
The ideal {\em random}
physical tendency 
${\bf P}_{\rm ideal}^*({\bf x}_{\rm GS})$ 
is a `possible true $\boldsymbol\pi$
consistent with the grid-scale field ${\bf x}_{\rm GS}$', that is, 
a random draw from the probability distribution $p(\boldsymbol\pi \mid {\bf x}_{\rm GS})$.
 It is reasonable to anticipate that
the aleatory part of the uncertainty, being caused by random subgrid noise, 
should be associated with small spatial and time scales.
On the contrary, the epistemic part of the uncertainty caused by 
systematic and, likely, flow-dependent deficiencies of the physical schemes,
may be characterized by larger spatio-temporal scales. 

In an ensemble, representing both kinds of uncertainty
in physical tendency using pseudo-random spatial fields results 
in {\em model perturbations} introduced at each model time step
during the forecast.
Techniques that add stochasticity on top of existing deterministic physical parametrization schemes
or replace the deterministic schemes with stochastic ones are known as `stochastic physics'.

\subsection{Existing stochastic physics schemes}
\label{sec_exist}

The first question is how to represent the {\em aleatory} uncertainty in physical tendency in ensemble
prediction schemes.
We believe that the most sensible way to address this question is to build intrinsically 
{\em stochastic} (rather than traditional deterministic) parametrization schemes.
Research in this direction is underway, see \citet{PlantCraig,Dorrestijn,Sakradzija,HirtCraig,Machulskaya,Clark}
and others.
The second question is how to represent the {\em epistemic} uncertainty in physical tendency
produced either by a deterministic or a stochastic physical parametrization scheme.

We state that, on the one hand, stochastic parametrizations have 
not yet replaced deterministic ones.
So we still need techniques to represent uncertainties in 
{\em deterministic} physical parametrizations.
Ad-hoc approaches are in wide use here.
On the other hand, stochastic parametrizations are not going to be devoid of epistemic 
uncertainties either. To represent those uncertainties, we  will, most likely, resort to ad-hoc schemes, too.
Our focus in this research is on ad-hoc model perturbation schemes.

Currently, the two most popular ad-hoc techniques to represent uncertainties in physical parametrizations
are Stochastically Perturbed Parametrization Tendencies
\citep[SPPT, ][]{Buizza,Leutbecher}
and Stochastically Perturbed  Parametrizations \citep[SPP,][]{Ollinaho}.
SPPT generates model perturbations relying on the assumption that 
the magnitude of the  error in the physical tendency
is proportional to the magnitude of the physical tendency itself.
There is also a flavor of SPPT called iSPPT in which tendencies from different physical
parametrizations are perturbed independently \citep{iSPPT}.
SPPT has proven to be very useful in practical ensemble prediction schemes 
despite its lack of physical consistency: it does not respect
conservation laws because it perturbs tendencies without modifying fluxes, see \citet{Leutbecher},
\citet{LangSPP}, and  references therein.

In SPP, selected parametric and structural elements of the model's parametrization schemes 
are made spatio-temporal random fields rather than fixed numbers and fixed choices.
Advantages of SPP include 
(i) reliance on expert knowledge to select perturbed  elements and 
design probability distributions, 
(ii) internal consistency of the resulting numerical scheme and conservation properties, and 
(iii) capability of generating significant spread in the ensemble \citep{LangSPP,SPP_canada2}.
The disadvantages of SPP are  more conceptual. 
First, it accounts only for uncertainties that can be captured by perturbing the
specific parametrization schemes used in the forecast model in question. 
Given the inevitably simplified nature of many parametrization schemes,
the resulting tendency perturbations may not explore some relevant directions in phase space.
Second, it is hard to even suggest how, say, 
parameter-perturbation probability distributions can be objectively justified. 
The reason is that parameters (and structural elements) of parametrization schemes may have no counterparts in nature 
(there is no `diffusion coefficient' in nature) and even in a high-resolution model 
used as a proxy to the truth. 
Besides, both SPPT and SPP can lead to biases, see, \eg \citet{Leutbecher} and \citet{Bouttier2022}, respectively.

We selected SPPT (described in more detail in section \ref{sec_SPPT}) as a starting point 
for our development in this study
because it attempts to do exactly what is needed to represent uncertainty
in physical parametrizations: it perturbs
the physical tendency (see above section \ref{sec_ErrUncert}).

In this study, we analyze limitations of  SPPT  and build a technique that attempts to address
those limitations. The new scheme termed 
Additive Model-uncertainty perturbations scaled by Physical Tendencies (AMPT) 
is  tested in numerical experiments with a convective-scale ensemble prediction system.

\section{Methodology}
\label{set_meth}

In this section,
we review SPPT  and introduce a new approach  to  generation of model-uncertainty perturbations.
The new AMPT scheme builds on SPPT and attempts to avoid/relax some of the deficiencies of SPPT.
AMPT is applied to both atmosphere and soil.

\subsection{Background on SPPT and notation}
\label{sec_SPPT}

To facilitate the presentation of   AMPT, we first outline  SPPT
\citep[][]{Leutbecher}. 
By ${\bf P}(x,y,\zeta,t) = (P_1(x,y,\zeta,t),\dots,P_{n_{\rm fields}}(x,y,\zeta,t))$ 
we denote the vector-valued physical tendency
(the net physical tendency, that is, generated by all physical parametrizations combined)
 at the spatial grid point with the Cartesian horizontal coordinates $(x,y)$,
the vertical coordinate $\zeta$, and the forecast time  $t$.
Here $P_i(x,y,\zeta,t)$ is the component of ${\bf P}(x,y,\zeta,t)$  in the $i$-th 
model field (variable) $X_i$, and
 $n_{\rm fields}$ is the number of model fields selected to be perturbed
\citep[most often, temperature, winds, and humidity, e.g.,][]{iSPPT}.

In SPPT, the perturbed physical tendency ${\bf P}^*$  is postulated to be
\begin {equation}
\label{SPPT}
{\bf P}^*(x,y,\zeta,t) = (1+\kappa\,\xi(x,y,t)) \, {\bf P}(x,y,\zeta,t),
\end {equation}
%
where $\xi(x,y,t)$ is the zero-mean and unit-variance spatio-temporal random field 
and $\kappa$ the  {\em scalar} (i.e. the same for all model variables) parameter 
that controls the magnitude of the perturbation. 
For stability reasons \citep[to avoid  sign reversal 
of the physical tendency, see, e.g.,][]{Leutbecher}, the support of the probability distribution of 
${\kappa}\,\xi(x,y,t)$ is limited to the segment $[-1,1]$ so that
\begin {equation}
\label{sign_rev}
|\xi(x,y,t)| < {\kappa}^{-1}.
\end {equation}
This constraint  limits the magnitude of perturbations in SPPT.

From Eq.(\ref{SPPT}), the   perturbation of the physical tendency in SPPT is seen to be multiplicative
with respect  to the model physical tendency:
\begin {equation}
\label{d_SPPT}
\Delta{\bf P} = {\bf P}^*(x,y,\zeta,t) - {\bf P}(x,y,\zeta,t) =  \kappa\, \xi(x,y,t)  \,{\bf P}(x,y,\zeta,t).
\end {equation}
Here and elsewhere $\Delta$ denotes a perturbation.

\subsection{Motivation}
\label{sec_motiv}

The following deficiencies of SPPT led us to propose the new approach. 

\begin{enumerate}
\item
\label{list_zero_tend}

In SPPT, perturbations are large (small) when and where  the
physical parametrizations generate a large  (small) physical tendency 
${\bf P}$.
This formulation gives rise to a meaningful scaling of perturbations in 
situations when the model predicts a high or moderate intensity of subgrid-scale processes 
and produces a large or moderate physical tendency.
However, it cannot cover situations
in which  the physical tendency appears to be small
or even zero whereas
the model uncertainty  is, in fact, large.
This may occur if, for example, in a grid cell, convection is initiated in nature
whilst the convective parametrization fails to be activated (note that in this case
switching to iSPPT would not help either).

\item
 Equation (\ref{d_SPPT}) implies that the multivariate perturbation vector
$\Delta{\bf P}(x,y,\zeta,t)$
is  strictly proportional 
 to the physical tendency vector ${\bf P}(x,y,\zeta,t)$. 
 As noted by \citet{Leutbecher}, this implies that SPPT
  tacitly assumes that  only the {\em magnitude}
 of the vector ${\bf P}$ is in error and not its direction, which is highly unlikely. 
In other words,  SPPT `assumes' that the {\em ratios}  of the physical tendencies in different variables
$i$ and $j$ at  the same  point $(x,y,\zeta,t)$ are {\em error-free}.
%
As a consequence, the SPPT perturbations (and thus the assumed model uncertainties) 
are perfectly (100\%) correlated or perfectly (-100\%) anticorrelated for any pair of model 
variables, which is not realistic.

Indeed, from Eq.(\ref{d_SPPT}) written component-wise, we have  $\Delta P_i=\kappa\,\xi P_i$, 
where $\kappa$ is a positive constant and $\Ex\xi=0$.
This implies that 
$\Ex (\Delta P_i \mid P_i) =0$, 
and 
$\Cov(\Delta P_i, \Delta P_j \mid P_i, P_j ) = \Ex ( \Delta P_i  \Delta P_j \mid P_i, P_j ) = \kappa^2 (\sd\xi)^2 P_i P_j$
(where $\Cov$ stands for covariance).
Taking into account that $\sd (\Delta P_i  \mid P_i )= \kappa \, |P_i|\,\sd\xi$ and
$\sd (\Delta P_j  \mid P_j )= \kappa \, |P_j|\,\sd\xi$, 
the correlation, that is, the  covariance normalized by the product of the two standard deviations, becomes
 $\Corr(\Delta P_i, \Delta P_j \mid P_i, P_j ) = \pm 1$.

\item
\label{list_vert}
Similarly \citep[and also noted by][]{Leutbecher}, since 
the  SPPT random pattern $\xi(x,y,t)$ does not depend on the vertical coordinate,
the  SPPT perturbations are perfectly coherent (correlated) 
for all variables at all levels in a vertical column, which again is unrealistic.

\item
\label{list_500km}
Moreover, it has appeared that for SPPT to give rise to a significant spread in the ensemble,
the length and time scales need to be really large for convective-scale models.
\eeg in \citet{Walser}, the tuned length and time scales in the 
2.2-km resolution COSMO model were as large as 500 km and 6 h, 
respectively.
This implies that the above unphysical $\pm 100$\% correlation of SPPT perturbations
approximately holds for all model variables in huge 4D volumes spanning the whole atmosphere in the vertical,
hundreds of kilometers in the horizontal, and hours of forecast time.

\item
\label{list_underdisp}
In some models, with their specific configurations of model perturbation schemes,
SPPT appeared to generate not enough spread in ensemble forecasts,
preventing them from producing reliable probabilistic forecasts,
see e.g. \citet{iSPPT,Frogner}.
One possible reason for that is the limitation on the magnitude of perturbation, Eq.(\ref{sign_rev}).

\end{enumerate}

With iSPPT \citep{iSPPT}, the above points \ref{list_zero_tend}--\ref{list_vert} pertain to the tendency due to 
a single parametrization scheme. iSPPT alleviates  weak points \ref{list_500km}--\ref{list_underdisp},
allowing for smaller spatial scales of the random pattern and  
having the capability to generate a somewhat larger
spread in the ensemble than SPPT \citep{Wastl,iSPPT}.

For completeness, it is worth mentioning that in developing AMPT we did {\em not} aim at alleviating
SPPT's lack of physical consistency.

\subsection{Approach}
\label{sec_approach}

In AMPT, we propose to address the  above deficiencies of SPPT as follows.

\subsubsection{Univariate AMPT design}
\label{sec_univar}

We rely on the SPPT's assumption that the standard deviation  
of the model-uncertainty perturbation $\Delta X_i = \Delta P_i$  is proportional to the modulus of 
its respective physical tendency, 
$\sd(\Delta X_i  \mid P_i) = \kappa |P_i|$.
In AMPT, we define this dependency to be more general than just point-wise,
postulating $\sd(\Delta X_i \mid P_i)$ to be proportional to 
an {\em area-averaged} $|P_i|$.
This allows AMPT to generate non-zero perturbations even at grid points with zero physical tendency ---
if there are nearby points with non-zero physical tendency.
Theoretically, this approach can be justified as follows.

Consider the unknown true model-uncertainty field $\varepsilon_i({\bf r})$
(where ${\bf r}$ is the spatio-temporal coordinate vector $(x,y,\zeta,t)$).
Assume that  $\varepsilon_i({\bf r})$ can be modeled as a random field with zero mean and
the unknown spatially variable standard deviation $\sigma_i({\bf r})$.
In these settings, SPPT, effectively, estimates $\sigma_i({\bf r})$ as $\kappa |P_i({\bf r})|$,
see Eq.(\ref{d_SPPT}), that is, the only predictor to estimate $\sigma_i$ at the point ${\bf r}$ 
in space and time
is the physical tendency at the same point. 
We propose to acknowledge that $|P_i({\bf r})|$ is a {\em noisy} `observation' of 
the true model-uncertainty standard deviation $\sigma_i({\bf r})$ and therefore it is
worth looking for other predictors, \ie other data that can contain information on $\sigma_i({\bf r})$.
In AMPT, we hypothesize that these additional relevant data are values of the absolute physical 
tendency in the {\em vicinity} of the spatio-temporal  point in question ${\bf r}$.
Linearly combining these noisy data, we obtain 
the AMPT's estimate of the unknown $\sigma_i({\bf r})$ given the known field $|P_i(.)|$:
\begin {equation}
\label{Wi}
\widehat\sigma_i({\bf r})  = \int W_i({\bf r,r}') \, |P_i({\bf r}')| \,\d {\bf r}',
\end {equation}
where $W_i({\bf r,r}')$ is a weighting function 
that determines the contribution of the absolute physical tendency, $|P_i|$, evaluated at the grid point ${\bf r}'$
to the estimate of the model-error standard deviation, $\widehat\sigma_i$, evaluated at the grid point ${\bf r}$
and the integral is over the model domain (and, possibly, over model time as well).
With the simplifying assumption that the weighting function is homogeneous, $W_i({\bf r,r}') =  w_i({\bf r-r}')$,
we rewrite Eq.(\ref{Wi}) as
\begin {equation}
\label{kappa_wi}
\widehat\sigma_i({\bf r})  =  \kappa_i {\cal P}_i({\bf r}), 
\end {equation}
where $\kappa_i = \int w_i({\bf r}) \d {\bf r}$,
\begin {equation}
\label{calPi}
{\cal P}_i({\bf r}) =  \int K_i({\bf r-r}') \; |P_i({\bf r}')| \,\d {\bf r}'
\end {equation}
is the {\em local-area averaged} absolute physical tendency 
(we will call it the {\em scaling  physical tendency}), and
$K_i({\bf r}) = \frac{1}{\kappa_i} w_i({\bf r})$ is the averaging kernel  such that 
$\int K_i({\bf r}) \,\d {\bf r} = 1$.

Having estimated $\sigma_i({\bf r})$, we can simulate the model uncertainty as
\begin {equation}
\label{d_AMPT}
\Delta X_i({\bf r}) = \widehat\sigma_i({\bf r}) \, \xi_i({\bf r}) =
   \kappa_i \,{\cal P}_i({\bf r}) \, \xi_i({\bf r}),
\end {equation}
where $\xi_i({\bf r})$ is a  zero-mean and unit-variance Gaussian 
random field postulated to be  stationary in space and time.
Its spatio-temporal correlations are discussed below in section \ref{sec_len}.
Spatial (and temporal) non-stationarity of the model-uncertainty field $\varepsilon_i$ 
defined by Eq.(\ref{d_AMPT})
comes from 
variability in the  
scaling  physical tendency ${\cal P}_i({\bf r})$. Non-Gaussianity of the model-uncertainty field
 comes from 
the randomness of ${\cal P}_i({\bf r})$.


\subsubsection{Scaling physical tendency}
\label{sec_scaling}

Technically, with the gridded fields, the integral in Eq.(\ref{calPi})
is replaced with a sum. 
Given the layered structure of the atmosphere,
we perform the averaging in the horizontal only:
\begin {equation}
\label{dP_w}
{\cal P}_i(x,y,\zeta,t)= \sum_{q} \sum_{r} w_{qr} \left| P_i(x_{q},y_{r},\zeta,t) \right|,
\end {equation}
where 
$(x_{q},y_{r})$ is the horizontal grid point, 
$i$  (we recall) labels the model variable, and  $w_{qr}$ are
the averaging weights. The latter are specified to be non-zero only inside the averaging area
$|x_{q} - x| < A_i, |y_{q} - y| < A_i$, where 
$A_i$ is the half-size of the averaging area 
(the averaging length scale) 
in both $x$ and $y$ directions.
For simplicity and due to a lack of knowledge on the spatial structure of uncertainties
associated with $\left| P_i(x_{q},y_{r},\zeta,t) \right|$ as `observations' on 
$\sigma_i(x,y,\zeta,t)$ we adopt the simplest design:
the weights $w_{qr}$ are equal to each other within the averaging area and normalized so that for any $x,y$, we have
$\sum_{qr} w_{qr} =1$. 
In the context of limited area modeling, if the point $(x,y)$, 
where ${\cal P}_i$ is evaluated, is near the model's boundary,
$A_i$ is reduced so that the averaging area is still a square and is within the model domain.

If $A_i$ is greater than the size of the domain, the averaging is performed over the whole domain
so that ${\cal P}_i(x,y,\zeta,t)={\cal P}_i(\zeta,t)$ is the same for all grid points in the horizontal.
This is how we computed 
the scaling physical tendency
for atmospheric temperature, winds, and soil temperature.
For less Gaussian fields such as humidity and soil moisture, ${\cal P}_i$ is computed 
by averaging over a significantly smaller moving window centered at the grid point in question
(see section \ref{sec_expm} for details).
It is worth noting that dependence of ${\cal P}_i(\zeta,t)$ on model time $t$ is essential.
The reason is that during the forecast, the mean magnitude of physical tendency may undergo significant  
variations due to a passage of a front or a convective system, changes in convection, \etc

\subsubsection{Multivariate and 3D aspects}
\label{sec_multivar}

First, we allow in AMPT not only for errors in the {\em modulus} of the vector ${\bf P}$ but also for errors in the 
direction of ${\bf P}$.
We do so by introducing {\em independent} driving random fields $\xi_i$ for different model variables $X_i$
(see Eq.(\ref{d_AMPT})).
Though purely uncorrelated perturbations are unphysical, we rely on the forecast model to 
  introduce physically meaningful relationships between the variables during its adaptation to the
model perturbations. Since the magnitude of the perturbation at each model time step is small
compared with the natural variability\footnote{In the experiments described below, 
the mean absolute magnitudes of AMPT perturbations per time step were 
 as follows (at 8 a.m. local time).
Temperature: somewhat less than 0.001 K in the lower 3-km tropospheric layer and much less above it.
A horizontal wind component: from 0.015 m/s  near the ground 
to 0.0001 m/s at 3 km height and even less above 3 km.},
the adaptation is expected to go smoothly.

Second, to get rid of the perfect coherency of the perturbations in the vertical, we switch from the
3D random pattern
$\xi(x,y,t)$ in SPPT  to the
4D random fields $\xi_i(x,y,\zeta,t)$ in AMPT.

Third, the perturbation-magnitude multiplier $\kappa_i$ is variable-specific in AMPT.

\subsubsection{Length scales and perturbation magnitudes}
\label{sec_len}

Let us assess the horizontal length scale of the SPPT perturbation field, see Eq.(\ref{d_SPPT}).
To this end, with fixed $\zeta$ and $t$, 
suppose that the physical tendency $P_i(x,y,\zeta,t)$ is a stationary (homogeneous) 
and isotropic random field as a function of $x,y$.
Let its correlation function be denoted as $C_P(r)$, where $r$ is the horizontal distance.
Since the random pattern $\xi(x,y,t)$ is stationary, isotropic, and independent of $P_i$ by construction, 
the horizontal correlation function of the SPPT perturbation field $\Delta^{\rm SPPT}$ is 
$C^{\rm SPPT}_\Delta(r) = C_\xi(r) \, C_P(r)$, where $C_\xi(r)$ is the correlation function of  $\xi$.
Assuming, further, that $C_\xi(r)$ and $C_P(r)$ are twice differentiable, we obtain the {\em differential length scale} 
\citep[e.g.,][]{Monin_V2}
$L^{\rm SPPT}_\Delta$ of the perturbation field 
as $(L^{\rm SPPT}_\Delta)^{-2} = (C^{\rm SPPT}_\Delta)''(0)$.
Since  $C_\xi(0)=C_P(0)=1$ and $C_\xi'(0)=C_P'(0)=0$, we obtain
\begin {equation}
\label{LSPPT}
(L^{\rm SPPT}_\Delta)^{-2} = (L^{\rm SPPT}_\xi)^{-2} + L_P^{-2},
\end {equation}
where $L^{\rm SPPT}_\xi$ and $L_P$ are  the differential length scales of the random pattern $\xi$ 
and the physical tendency $P$, respectively.

As we noted in item \ref{list_500km} of section \ref{sec_motiv}, the optimally tuned
length scale of the SPPT random pattern in convective-scale applications appears to be much greater than
the respective scales of natural variability.
In terms of the differential length scales, we may therefore conclude that $L^{\rm SPPT}_\xi \gg L_P$. Then, 
Eq.(\ref{LSPPT}) implies that $L^{\rm SPPT}_\Delta \approx L_P$.
Since it is plausible that the optimally tuned  $L^{\rm SPPT}_\Delta$ approximates
{\em the unknown model-uncertainty length scale $L_\varepsilon$}, we obtain
\begin {equation}
\label{LeLP}
L_\varepsilon \approx L_P.
\end {equation}
In AMPT, the random pattern $\xi$ multiplies the  scaling physical tendency ${\cal P}$ so that
\begin {equation}
\label{LAMPT}
(L^{\rm AMPT}_\Delta)^{-2} = (L^{\rm AMPT}_\xi)^{-2} + L_{\cal P}^{-2},
\end {equation}
where $L_{\cal P}$ is the  differential length scale of ${\cal P}$.
Since ${\cal P}$ is a {\em spatially smoothed} version of $|P|$, 
$L_{\cal P} \gg L_P$.
Therefore, for the length scale of the AMPT perturbation to be close to $L_P$, 
we have to make $L^{\rm AMPT}_\xi \approx L_P$,   that is, much smaller than in SPPT.
The optimally tuned length scales of the AMPT random patterns are given in section \ref{sec_expm}
(they, indeed,  appeared to be an order of magnitude smaller than their SPPT counterparts).

As for the AMPT magnitude multipliers $\kappa_i$, the {\em non-local} dependence of the scaling physical tendency
on the modulus of the unperturbed physical tendency (Eqs. (\ref{calPi}) and (\ref{dP_w})) suggests that $\kappa_i$
do not need to obey the SPPT's strict upper limit
on the magnitude multiplier,   Eq.(\ref{sign_rev}).
This allows AMPT to cause greater spread than SPPT in the ensemble if needed.

\subsubsection{Further details}
\label{sec_det}

The rest of the Methodology section is organized as follows.
Details on  AMPT perturbations for specific model fields  are given below in section \ref{sec_specif}.
The Stochastic Pattern Generator 
 \citep[SPG,][]{TsyGaySPG} 
 is outlined in section \ref{sec_SPG}.
Section \ref{sec_map} explains how  SPG fields are mapped from the SPG domain onto the model domain.
In section \ref{sec_prop} we briefly discuss stability, conservation properties of
the new scheme,  possible biases due to nonlinearity of the forecast model, 
and explain the terminology according to which we call SPPT perturbations multiplicative and
AMPT perturbations additive.

\subsection{Treatment of specific model fields}
\label{sec_specif}

In the atmosphere, we experimented with perturbations of  the
3D fields of temperature $T$, pressure $p$, horizontal wind components $u,v$, and
specific humidity $q_{\rm v}$. 
We also tried perturbing  cloud ice and cloud water but found that those perturbations 
had little overall impact, so we abandoned them.
In the soil, we perturbed 3D fields of soil temperature and
soil moisture.

\subsubsection{Atmospheric temperature, pressure, and winds}
\label{sec_atm}

Independent perturbations of $T,u,v$ are computed following Eq.(\ref{d_AMPT}).
The pressure perturbation $\Delta p$ is computed
from $\Delta T$  by
integrating the hydrostatic equation (in which  $\Delta q_{\rm v}$ is neglected
as a small contribution to a small perturbation) assuming zero pressure perturbation at the top
of the model domain. 

As for wind perturbations, we note that theoretically, it is `better' to rely on
mutually uncorrelated random {\em stream function} and 
{\em velocity potential} perturbation fields ---
rather than on  mutually uncorrelated $u$ and $v$ (\ie zonal and meridional wind) perturbation fields.
The reason is that the former approach allows for {\em isotropic} vector-wind perturbations 
\citep[][section 12.3]{Monin_V2}, unlike the latter approach.
However, in practical terms, we were not able to identify any practically significant flaw
in the vector field composed of two independent $u$ and $v$ perturbation fields.
For this reason and due to the  lack of evidence on the actual structure of model uncertainties, 
we stick to the simpler formulation of AMPT with mutually
independent $\Delta u(x,y,\zeta,t)$ and $\Delta v(x,y,\zeta,t)$ 
in this study.


\subsubsection{Humidity}
\label{sec_qv}

The salient difference of  humidity $q_{\rm v}$ from $T,u,v$ is that $q_{\rm v}$ has a 
narrow range of values (from zero to saturation or somewhat higher than saturation).
The range of  $q_{\rm v}$ is narrow  in the sense that it is comparable to the standard deviation of 
the natural variability in $q_{\rm v}$.
To make sure that the AMPT-perturbed $q_{\rm v}$ is within this range (\ie from zero to saturation) and
does not  directly introduce any {\em bias} into the model, we modify the above formulation of AMPT.
Specifically, we employ  a kind of `perturbation symmetrization'  as follows.
Consider a grid point ${\bf s}$ and model time  $t$, where and when the unperturbed specific humidity is $q_{\rm v}({\bf s},t)$.
We compute the tentative perturbation $\Delta q_{\rm v}({\bf s},t)$  following Eq.(\ref{d_AMPT})
and then symmetrically {\em truncate} it at $\pm c$, where
$c = \min(q_{\rm v}({\bf s},t),\, q_{\rm sat}({\bf s},t) - q_{\rm v}({\bf s},t))$
and $q_{\rm sat}({\bf s},t)$ is the saturated specific humidity.
Due to the truncation,  the perturbed 
$q_{\rm v}^*({\bf s},t) = q_{\rm v}({\bf s},t) + \Delta q_{\rm v}({\bf s},t)$ is, first, 
within the admissible range from 0 to $q_{\rm sat}({\bf s},t)$ and second, 
as the truncation is symmetric, no bias is directly introduced:
$\Ex (q_{\rm v}^*({\bf s},t) \mid q_{\rm v}({\bf s},t) ) = q_{\rm v}({\bf s},t)$ 
(which would not be the case if we just truncated $q_{\rm v}^*({\bf s},t)$ at 0 and 
$q_{\rm sat}({\bf s},t)$).

\subsubsection{Soil fields}
\label{sec_soil}

In the land (soil) model, tendencies of two model fields are perturbed: soil temperature  $T_{\rm so}$ and
soil moisture (more specifically, soil water content $W_{\rm so}$ per unit area within the soil layer in question).
Compared with the atmospheric AMPT, the differences in 
the treatment of the soil fields are the following.
\begin{enumerate}
\item
The  scaling tendencies  ${\cal P}_{T_{\rm so}}$ and ${\cal P}_{W_{\rm so}}$
are computed by averaging  the total  tendency, 
assuming that   all processes  in the soil are modeled with substantial uncertainty.

\item
The averaging in Eq.(\ref{dP_w}) is performed over land only.

\item
The perturbation patterns in the soil,
$\xi_{T_{\rm so}}(x,y,t)$ and $\xi_{W_{\rm so}}(x,y,t)$, 
used to generate perturbations following Eq.(\ref{d_AMPT})
are three-dimensional
(not four-dimensional as in the atmosphere).

\end{enumerate}
%


 {\em Whole-domain} averaging 
 is used to compute the scaling physical tendency ${\cal P}_{T_{\rm so}}$.
{\em Local-area} averaging (\ie with the averaging length scale $A_{W_{\rm so}}$ much smaller than the domain size)
is employed  to  compute the scaling physical tendency  ${\cal P}_{W_{\rm so}}$.
This choice is motivated by higher variability/non-Gaussianity of soil moisture compared
to soil temperature (not shown).

The perturbed $W_{\rm so}$ is truncated to ensure 
that the volumetric soil water content 
$\eta_{\rm so}$
is between the wilting-point $\eta_{\rm wp}$ and 
field capacity $\eta_{\rm fc}$:
\begin {equation}
\label{eta_SO}
\eta_{\rm wp} \le \eta_{\rm so} \equiv \frac{W_{\rm so}(x,y,Z,t)}{\Delta_Z} \le \eta_{\rm fc},
\end {equation}
where $Z=1,2,\dots,n_Z$ labels the soil layer, $\Delta_Z$ is the thickness of the $Z$-th layer,
and  $n_{\rm Z}$ is the number of layers.

\subsubsection{Initial soil perturbations}
\label{sec_soil_ini}

In the soil,  processes have much longer time scales than in the atmosphere, therefore the role
of model perturbations can be revealed only in long-range forecasts or in cycled systems.
Dealing in this study with short-range forecasts without cycling and having an under-dispersive ensemble of initial
conditions, we had to develop 
a generator of {\em initial} $T_{\rm so}$ and $W_{\rm so}$  perturbations.

The initial soil temperature perturbation is specified as
\begin {equation}
\label{dTsoil_ini}
\Delta T_{\rm so}^{\rm ini}(x,y,Z) = \kappa^{\rm ini}_{T_{\rm so}} \, c_{T_{\rm so}}^{-Z} \, \xi(x,y), 
\end {equation}
where $\kappa^{\rm ini}_{T_{\rm so}}$ is the external magnitude parameter, $c_{T_{\rm so}}>1$ is the 
vertical-decay external parameter, 
$Z=0,1,\dots, n_{\rm Z}$ (with 
$Z=0$ denoting the so-called surface temperature and $Z=1,\dots, n_{\rm Z}$
labeling the  soil layers), and
$\xi(x,y)$ is the 2D pseudo-random field.
Note that the random pattern $\xi$ is the same for all soil layers,
whilst the magnitude of the perturbation exponentially decreases downwards.

With the soil moisture $W_{\rm so}$, the technique  \citep[inspired by][]{Schraff} is to  perturb the 
Soil Moisture Index 
\begin {equation}
\label{S}
S = \frac{\eta_{\rm so} - \eta_{\rm wp}} {\eta_{\rm fc} - \eta_{\rm wp}}
\end {equation}
as follows:
\begin {equation}
\label{Sxyz}
\Delta S(x,y,Z) = \kappa^{\rm ini}_{S} \, c_{W_{\rm so}}^{1-Z}  \, \xi(x,y),
\end {equation}
where $ \kappa^{\rm ini}_{S}$ is the magnitude parameter,
$c_{W_{\rm so}}$ is the vertical-decay parameter, and $Z=1,2, \dots, n_{\rm Z}$.
If at a grid point, the perturbed $S$ appears to lie outside the meaningful range $[0,1]$, 
the perturbation $\Delta S$ is truncated accordingly. 
Using Eqs. (\ref{S}) and (\ref{eta_SO}), we finally convert the perturbation $\Delta S(x,y,Z)$ into the 
perturbation of  $W_{\rm so}$:
\begin {equation}
\label{Wpert}
\Delta W_{\rm so}^{\rm ini}(x,y,Z) =  \Delta_Z \, [ \eta_{\rm wp} + 
   \Delta S(x,y,Z)  (\eta_{\rm fc} - \eta_{\rm wp})].
\end {equation}
%

\subsection{Stochastic pattern generator (SPG)}
\label{sec_SPG}

In this study, we rely on the limited-area Stochastic Pattern Generator (SPG) developed by \citep[][]{TsyGaySPG}
to generate independent four-dimensional spatio-temporal pseudo-random fields $\xi_i$ needed by AMPT, see
Eq.(\ref{d_AMPT}). Each $\xi_i$ is computed by solving the stochastic pseudo-differential equation
\begin {equation}
\label {SPG}
 \left( \frac {\partial }{\partial t} +  \frac{U}{\lambda} \sqrt{1 - \lambda^2\nabla^2} \right)^3  \xi(t,x,y,z) =
  \sigma \alpha(t,x,y,z).
\end {equation}
Here $x,y,z$ are the SPG-space spatial coordinates,
$\nabla^2$ is the 3D Laplacian,
$\alpha(t,x,y,z)$ is the Gaussian  white  noise, and $\lambda, U, \sigma$ are the parameters.
$\lambda$  determines the spatial scale of $\xi$. 
Given $\lambda$, the characteristic velocity $U$ determines the time scale  of $\xi$. 
Given $\lambda$ and $U$, the parameter $\sigma$ determines the variance of $\xi$  and is selected to ensure that $\sd\xi=1$.
The computational domain is the  cube of size $2\pi$ and
periodic boundary conditions in all three dimensions.

The design of Eq.(\ref{SPG}) (note the third order of the equation in time and the square root of the negated and shifted Laplacian) 
is dictated by two requirements:
\begin{enumerate} 
\item
The solution $\xi$ satisfies
the so-called proportionality of scales property \citep{Tsyroulnikov2001}.
The meaning of this property is the following.
For any $t$, let us expand $\xi$ in Fourier series in space:
$\xi(t,x,y,z) = \sum \widetilde\xi_{m n \ell}(t) \,\e^{\i (mx +ny+\ell z)}$, where 
$m,n,\ell$ are the spatial wavenumbers and
$\widetilde\xi_{m n \ell}(t)$ are the Fourier coefficients. Then 
Eq.(\ref{SPG}) decouples into a series of ordinary stochastic differential equations for the random processes
$\widetilde\xi_{m n \ell}(t)$. The proportionality of scales
means that for a large total  wavenumber ${\cal K}=\sqrt{m^2+n^2+\ell^2}$,
the {\em time scale of the process $\widetilde\xi_{m n \ell}(t)$ is proportional to its
 spatial scale} $1/{\cal K}$, a property often possessed by natural spatio-temporal processes.
 
\item
The spatial spectra of $\xi$ should be convergent, that is, the process variance
should be finite in the space-continuous case and thus
should be bounded above as
the spatial resolution increases in the space-discrete case.

\end{enumerate}

The solution to Eq.(\ref{SPG}) is a zero-mean and unit-variance 
homogeneous (stationary in time and space) 4D random field, which 
has `nice' spatio-temporal correlations: the shape of the correlation function
is the same along any direction in the 4D space (thus, including spatial and temporal correlations).
This universal correlation function belongs to the very popular in spatial statistics Mat\'ern class,
see \citep[][]{TsyGaySPG} for details.

Since the SPG computational domain is the cube whereas the limited-area-model domain can
be approximated by a rectangular  parallelepiped, we  
use the anisotropic Laplacian $\nabla^2_*$ instead of $\nabla^2$ in Eq.(\ref{SPG}): 
\begin {equation}
\label {Lapl_anis}
\nabla^2_* = \frac{\partial^2}{\partial x^2} +  \gamma^2 \frac{\partial^2}{\partial y^2} +  
             \delta^2\frac{\partial^2}{\partial z^2}.
\end {equation}
Here $\gamma$ is the aspect ratio of the model domain in the horizontal and
$\delta$ controls the length scale along the  $z$ axis. 

The SPG internal parameters (for the model variable labeled by $i$) $\lambda_i$ and $\delta_i$  are computed from 
the respective horizontal and vertical length scales, $L_i$ and $H_i$,
which are SPG external parameters
(this is done by taking advantage of the known correlation functions of the 4D SPG field in space and time).
The time scale is ${\cal T}_i = L_i /U$, where $U$ is the above characteristic velocity (another
external parameter).

With each realization of the driving white noise $\alpha$, 
the solution  to Eq.(\ref{SPG}) is computed in 3D Fourier space 
using the backward Euler finite difference scheme in time.
Once per ${\cal T}^{\rm FFT}$ model time, the solution is
converted to physical space (using the 3D inverse fast Fourier transform, FFT).
The resulting physical-space random pattern $\xi(t,x,y,z)$ is then mapped onto the model domain
and used to compute model perturbations (according to Eq.(\ref{d_AMPT}))
for the next ${\cal T}^{\rm FFT}$ model time interval.
Avoiding the application of the inverse Fourier transform every time step
significantly reduces the computational cost of the numerical scheme.
${\cal T}^{\rm FFT}$ is selected to be a few times less than the time scale of the process,
${\cal T}_i$.

SPG can also generate 3D fields $\xi(t,x,y)$ and $\xi(x,y,z)$ as well as 2D fields $\xi(x,y)$.

\subsection{Mapping fields from SPG grid to model grid}
\label{sec_map}

In each of the three spatial dimensions, to avoid significant unphysical 
correlations in $\xi_i$ between the opposite sides of the model domain due to the periodicity of the SPG domain, 
a segment of length $2 L_i$ in the SPG domain is, first, discarded. Then,
the rest of the SPG domain (the `working domain') is  mapped onto the model domain, where the field is 
interpolated to the model grid points.
In the horizontal, the mapping is piecewise linear.
In the vertical, two options were explored. 

With Option 1, the mapping $z \mapsto \zeta$ 
(where $\zeta$ is the model's vertical coordinate) 
is piecewise linear (just as in the horizontal) 
so that in the model space, the resulting field
is stationary (homogeneous) in the vertical as a function of $\zeta$.

With Option 2, a non-stationarity (inhomogeneity) in the vertical is introduced.
Specifically, we assume that the model vertical levels are unevenly positioned by the model designer 
to account for the variable vertical length scale of the model fields.
Say, in the planetary boundary layer, the model levels are dense, reflecting shallower 
meteorological structures (and thus shorter vertical length scales) 
than in the troposphere and stratosphere, 
where the vertical grid is much coarser because the vertical length scales are greater.

To obtain the same vertical inhomogeneity in AMPT perturbations interpolated to the model grid, 
we employ a mapping that is linear from the SPG coordinate $z$ to 
the model vertical (continuous) `computational' coordinate
defined to be equal to $\ell$ at the model level $\ell$.
With this mapping, 
the correlations between all adjacent model grid points in the vertical are the  same
(due to stationarity of the SPG field as a function of $z$).
As a result, in  model space, the vertical correlation falls to the same value at short
vertical distances in the  boundary layer (because the model grid is dense there)
and at longer vertical distances above the boundary layer (where the model grid is coarser)
thus implying the desired inhomogeneity in the vertical.

\subsection{Properties of AMPT}
\label{sec_prop}

\subsubsection{Stability}
\label{sec_stabil}

Stochastic-dynamic systems in which the magnitude of random forcing depends on the current
state of the system and the forcing (perturbation) is time-correlated, 
can be unstable due to a positive feedback loop.
Indeed, a deviation of the model state  from its 
mean value may result in a greater forcing, which may lead to an even greater
deviation of the state from the mean, and so on until `explosion'.
To break this vicious circle,  we considered two strategies.

A technically simpler one, which we adopted in this study, is to update the scaling 
tendency ${\cal P}_i$ not at every time step 
but less frequently,
once per ${\cal T}_{{\cal P}_i}^{\rm update}$ model time 
during the forecast, where ${\cal T}_{{\cal P}_i}^{\rm update}$
is an external parameter defined below in section \ref{sec_expm}.

A somewhat more involved but potentially more powerful approach (which we left  for future research)
is to calculate the scaling physical tendency from the {\em unperturbed} (control) model run, ${\bf x}^{\rm control}$,
let us denote it  ${\cal P}_i({\bf x}^{\rm control}({\bf s}, t))$ 
(where $i$ labels the model variable, ${\bf s}$ is the  spatial grid point, and $t$ is  time). 
The scaling physical tendency, being a moving average of the modulus of the physical tendency,
is a smooth field, so  ${\cal P}_i({\bf x}^{\rm control}({\bf s}, t))$ can be stored 
during the control run in a file  on a {\em coarse} spatio-temporal grid (meaning that this 
should be feasible in terms of required computer resources).
Then, ${\cal P}_i({\bf x}^{\rm control}({\bf s}, t))$ can be  used to compute AMPT perturbations for all ensemble members.
By construction, this device will  destroy the harmful positive feedback loop.
We remark that taking fields from the control run to make model
perturbations  state-dependent
can be used to prevent instabilities not only with AMPT but with other techniques as well.

However, as noted by an anonymous reviewer, this technique cannot be applied to long-range predictions
where ensemble members may diverge from the unperturbed member to the extent that 
the physical tendency produced by the latter becomes irrelevant for the former.
Another caveat is that an application of that technique to an {\em operational} ensemble
prediction system will require that the unperturbed member be computed somewhat earlier than the rest
of the ensemble. 

Another promising approach is to perturb {\em fluxes} instead of prognostic fields,
as discussed in section \ref{sec_cons}.

\subsubsection{Conservation properties}
\label{sec_cons}

Neither SPPT nor AMPT respects local conservation laws.
The reason is that in both SPPT and AMPT, the model fields are perturbed whereas the {\em fluxes} are not.
Switching from perturbing state variables to perturbing fluxes
\citep[as done by][for deep convection]{Ginderachter}, 
including boundary fluxes, would  solve the problem. 
%
%
However, a purely flux-based model perturbation scheme aimed to represent uncertainties
in multiple physical parametrizations remains to be built and tested, which is beyond the scope of this study.

\subsubsection{Biases}
\label{sec_biases}

It is well known and easy to understand  that feeding a {\em nonlinear} system with an unbiased signal can lead to a
biased system output. 
Biases can appear even in a linear system if it is perturbed in a  {\em multiplicative} way, like in SPPT.
\citet[][Figs. 3 and 5]{Bouttier2012} and \citet[][Fig. A1(a)]{Leutbecher}
found that in their systems, SPPT led to the drying of the atmosphere and reduced precipitation.
In experiments by \citet{Romine2014} the atmosphere also became too dry due to SPPT (their Figs. 5 and 8) but 
that was associated with an increase in precipitation (their Fig. 9).
One possible reason that led to different outcomes in those studies is the impact of the
supersaturation limiter (which nullifies temperature and humidity perturbations at the grid points where the water vapor is 
saturated or super-saturated).
We experimentally study forecast biases induced by various configurations of AMPT and compare them with forecast biases
induced by SPPT in section \ref{sec_hum_bias}.

\subsubsection{Additive \vs multiplicative model perturbations}
\label{sec_addMult}

In Appendix \ref{App_add_mult}, 
we discuss why SPPT perturbations are multiplicative and  why we call AMPT perturbations additive.
We show that  AMPT perturbations are truly additive in the setting
when the scaling physical tendency 
is taken from the control run  (this option was discussed in section \ref{sec_stabil}).

In numerical experiments presented below, the model perturbation scheme is an approximation 
to the setting in which the scaling physical tendency  is completely decoupled from the 
current model state. The coupling is only relaxed by updating the scaling physical tendency 
much less frequently than every model time step (see sections \ref{sec_stabil} and \ref{sec_expm}).
In this regime, AMPT perturbations are not multiplicative but not yet fully additive.
Nevertheless, with some abuse of terminology, we call them additive bearing in mind the above scheme
with the scaling physical tendency taken from the unperturbed model run.

\section{Experimental Settings}
\label{sec_expm}

The program code of  AMPT  was built into 
the limited-area non-hydrostatic COSMO  
model \citep[][]{Baldauf}, which has both atmospheric and soil prediction
modules. The model was, in turn, embedded in a limited-area ensemble prediction system.
The AMPT-generated model perturbations along with an ensemble of initial and lateral-boundary
conditions allowed us to compute (and verify) 
ensemble forecasts.
The goal was to assess the effect of AMPT on deterministic and probabilistic forecasts
and compare it with the effect of SPPT perturbations.

%

The COSMO model (version 5.01) was used in the  convection-permitting configuration 
with a horizontal grid spacing of 2.2 km, 172*132 grid points
in the horizontal, and 50  levels in the vertical. 
The model integration time  step was 20 s.
The model's vertical coordinate was the height-based hybrid (Gal-Chen) coordinate 
\citep{Gal-Chen}.

The ensemble prediction system used in this study was  COSMO-Ru2-EPS
\citep{Montani2014,Astakhova2015}, which was developed within the FROST-2014 international project \citep{Kiktev_FrostBAMS} 
and the CORSO priority project of the COnsortium for Small-scale MOdeling (COSMO) \citep{Rivin}.    
The ensemble size was 10. 
COSMO-Ru2-EPS performed a dynamical downscaling of the forecasts of the driving  COSMO-S14-EPS system developed by 
the Italian meteorological service ARPAE-SIMC 
\citep{Montani2014}.
Thus, both initial and lateral boundary conditions for the control forecast and  ensemble members
were provided by COSMO-S14-EPS, which had a horizontal grid spacing of 7 km and 40 vertical levels. 
COSMO-S14-EPS was a clone of the consortium ensemble prediction system COSMO-LEPS \citep{Montani2011}
with a smaller ensemble size. 

\begin{figure}[h]
\begin{center}
 \scalebox{0.6}{ \includegraphics{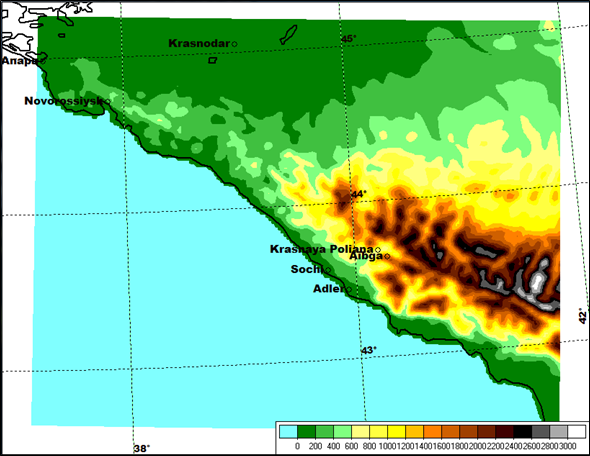}}
\end{center}
 \caption{Model domain and  orography.}
\label{fig_domain}
\end{figure}

The model domain  is shown in Fig. \ref{fig_domain}. 
Note the complexity of the area, which contains high mountains along with the adjoining valleys and  sea.
The center of the domain is located nearly at $44^\circ$N, $40^\circ$E. The climate at the sea level is   humid subtropical.
Numerical experiments were carried out in this study mostly for the winter-spring season: in February and March,
for which we had access to all data needed to run and verify ensemble forecasts.
Some sensitivity experiments were also conducted for May cases.

The  following AMPT parameters were selected for numerical experiments.
The horizontal length scale $L_i$ of the SPG driving random fields $\xi_i$ (where $i$
labels the perturbed fields)
was  tuned to be 50 km (\ie 20--25 horizontal mesh sizes) for $T,u,v$ and
35 km for $q_{\rm v}$. 
The vertical length scale $H_i$ was about 3 km for $T,u,v$ and about 2 km for humidity.

The characteristic velocity  $U=15$ m/s  was selected from the physically
meaningful interval of 10--20 m/s. 

After some experimentation, the (dimensionless)  model-uncertainty magnitude
multipliers $\kappa_i$ were specified as 0.75 for $T,u,v$, $T_{\rm so}$,  $W_{\rm so}$ and 
0.5 for  $q_{\rm v}$.

The length scale $A_i$ of averaging the absolute physical tendency  (see section \ref{sec_scaling}) 
was specified as equal to the length scale $L_i$ of the respective SPG
random field $\xi_i$ for $q_{\rm v}$ and $W_{\rm so}$. 
For  $T,u,v$ and $T_{\rm so}$, the respective $A_i$ were
selected large enough to ensure the whole-domain averaging of $|P_i|$.

The time update interval ${\cal T}_{{\cal P}_i}^{\rm update}$ of the scaling physical tendency (see section \ref{sec_stabil})
for the model field $X_i$ was set equal to the time scale ${\cal T}_i$ of the respective random pattern
$\xi_i$. Note that with ${\cal T}_{{\cal P}_i}^{\rm update}$
much less than ${\cal T}_i$, the perturbed model may become unstable (see section \ref{sec_stabil}), whereas
with  ${\cal T}_{{\cal P}_i}^{\rm update}$ much greater than ${\cal T}_i$,
the resulting area-averaged ${\cal P}_i$ may become irrelevant in a rapidly developing meteorological situation.

It is worth reiterating at this point that (i) we updated the scaling physical tendency once per 
 ${\cal T}_{{\cal P}_i}^{\rm update} \approx 1$ h model time, (ii) we computed the physical-space
 random patterns by performing the inverse Fourier transform of the spectral-space SPG fields
 once per ${\cal T}^{\rm FFT} = 20$ min model time, and (iii) we added model perturbations every model time step.

For the soil fields $T_{\rm so}$ and  $W_{\rm so}$, the common time scale  ${\cal T}_{\rm so}$ was 
specified 12 times as large as the atmospheric-temperature time scale ${\cal T}_{\rm T}$
(so the time scales of the perturbation fields were, roughly, 1 h in the atmosphere and 12 h in the soil).

In initial soil perturbations (see section \ref{sec_soil_ini}), 
the magnitude parameter  $\kappa^{\rm ini}_{T_{\rm so}}$, was 1 K.
The magnitude parameter of the initial  soil moisture index perturbation, $\kappa^{\rm ini}_{S}$
was only 0.01 (larger values led to unrealistically large model tendencies in $T_{\rm so}$).
The vertical-decay parameters were  $c_{T_{\rm so}}=1.75$  and $c_{W_{\rm so}}=2$.  

The mapping from the SPG space to the model space in the vertical was performed using Option 1
described in section \ref{sec_map} (for technical reasons we couldn't perform enough numerical 
experiments with the more physical Option 2 to judge which option is better).

In SPPT,
the spatial scale was about 500 km
and the time scale was 6 h. 
The random multiplier $\kappa\xi$ had the standard deviation 1 and then was truncated at the absolute value 0.8.
This SPPT setup implied greater perturbations than those explored with the same-resolution COSMO model by \citet{Walser}.
We opted for stronger SPPT perturbations because otherwise, they generated too little spread in the
ensemble forecasts.
The supersaturation limiter was off in SPPT.
In AMPT, the impact of the supersaturation limiter is
discussed below in section \ref{sec_superat}.

Tapering (\ie gradual reduction) of perturbations
(i) in the lower troposphere towards the surface and 
(ii) in the stratosphere from the tropopause upwards was handled in SPPT as follows.
The stratospheric tapering was always active  because it is believed that
the radiation tendency, which is dominant in the stratosphere, 
is quite accurate  in  clear-sky conditions   \citep[e.g.,][]{Leutbecher}.
As for the lower-tropospheric tapering, which is intended to prevent instabilities
due to  inconsistencies of perturbed physical tendencies and unperturbed surface fluxes
\citep{Wastl},
we found that SPPT worked better without it.
Specifically,  our experiments showed that, on the one hand, 
SPPT without tapering was stable in the boundary layer.
On the other hand, with tapering, SPPT led to an unacceptably small ensemble spread in the near-surface fields,
so we switched off the lower-tropospheric tapering in SPPT.
In AMPT, tapering was off everywhere.

Note that the above spatial and time scales employed in AMPT were an order of magnitude less than
those in SPPT (50 km \vs 500 km and 1 h \vs 6 h).
As discussed in section \ref{sec_len}, this is reasonable. 
If  in SPPT, $\xi(x,y,t)$ were small-scale, the product $\xi \, {\bf P}$ would become too patchy, 
reducing the effect of the perturbation on the forecast.
To find out if this argument is reasonable, we   ran SPPT with a smaller time scale of 1 h 
and  spatial scales of 50 km and 100 km.
The resulting spread in the ensemble forecast was indeed very small, confirming  the conclusions of \citet{Walser} 
(made for a different domain, orography, physiography, etc.) 
and justifying the choice of the SPPT  parameters for our domain
(the setup was also consistent with that employed in AROME-EPS by \citet{Bouttier2012}).


Ensemble forecasts were initialized every day at 00 UTC
during the two months of February--March, 2014. The local time was UTC+4h.

Our focus in this study was on near-surface fields: temperature, precipitation, and wind speed.

\section{Bias and spread induced by AMPT and SPPT}
\label{sec_hum_bias}

In preliminary experiments, we found that humidity perturbations  
did substantially increase ensemble spread but at the 
expense of introducing a significant bias into the forecast.
This led us to explore the impact of humidity perturbations (and the supersaturation limiter) 
on the spread of ensemble forecasts and the bias of the ensemble-mean forecast
(in terms of precipitation and near-surface temperature).
The aim was to decide whether it is worth perturbing humidity in the AMPT scheme at all.
The (informal) criterion was twofold: AMPT should generate significantly more spread in the ensemble forecast than SPPT
while having the bias-to-spread ratio as low as possible.
Note that by the  bias-to-spread ratio, we mean the absolute value of the bias divided by the spread.

In the experiments described in this section, to goal was to 
isolate the roles of different {\em model} perturbations. 
To this end, we set the same 
initial and lateral-boundary conditions for all ensemble members 
and switched off soil perturbations.
Additionally, hydrostatically balanced pressure perturbations in AMPT were deactivated.
The bias (caused solely by model perturbations) is defined in this section
as the domain-averaged difference between the ensemble mean  and the unperturbed deterministic forecast.

Results are presented in terms of bias-spread scatterplots for near-surface temperature and  accumulated total precipitation.
Bias and spread are combined for 
lead times from 3 h to 24 h (every three hours for precipitation and  every hour for 
near-surface temperature) and shown on a single scatterplot.
Each symbol on the plot represents the bias (the value on the $x$-axis) and the  spread
(the corresponding value on the $y$-axis) for a single lead time.

\subsection {Role of supersaturation limiter}
\label{sec_superat}

The question was how the supersaturation limiter impacts the spread of the ensemble
and the bias of the ensemble mean forecast.

\begin{figure}[h]
\begin{center}
 \scalebox{0.44}{ \includegraphics{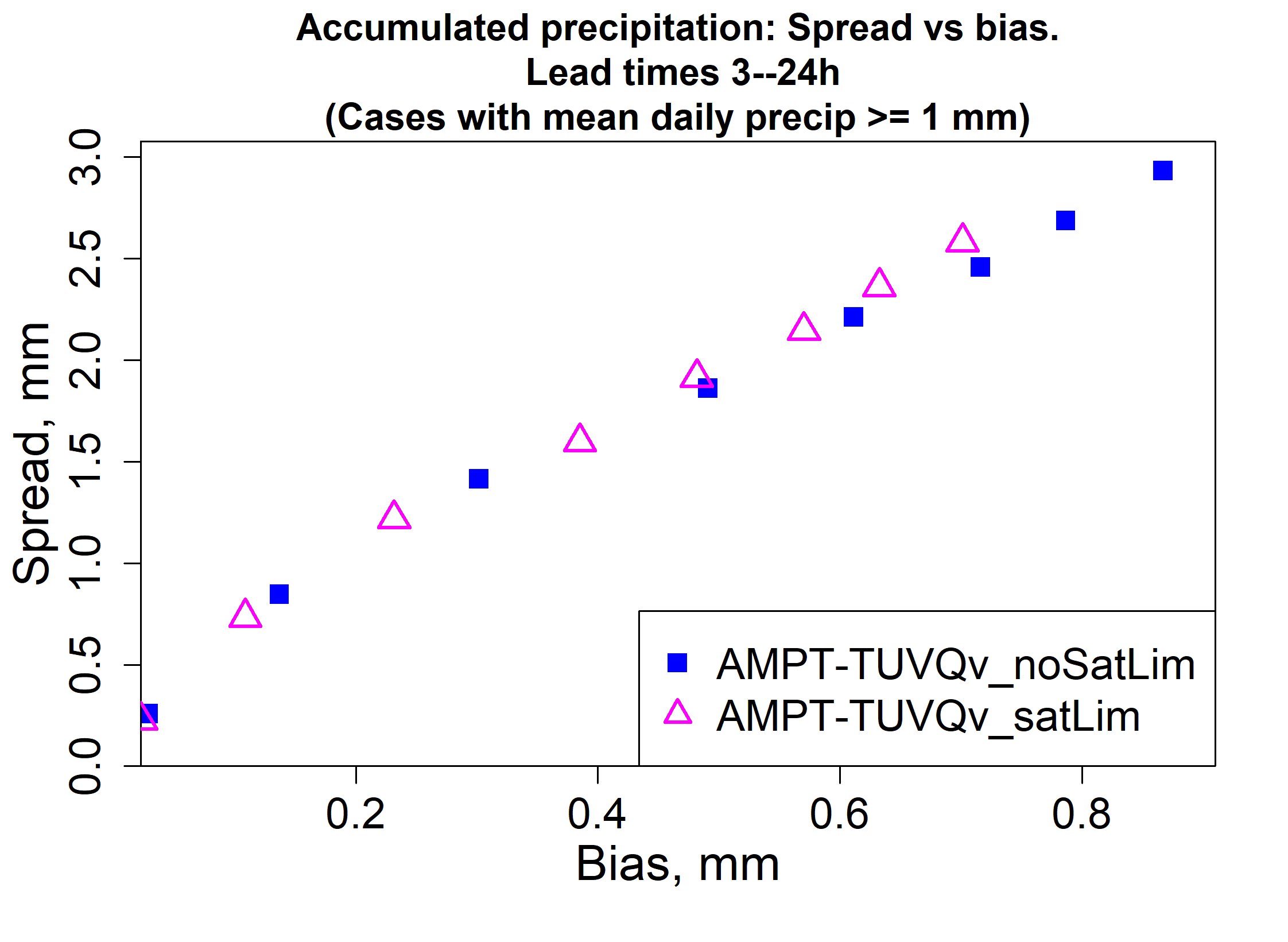}}
 \scalebox{0.44}{ \includegraphics{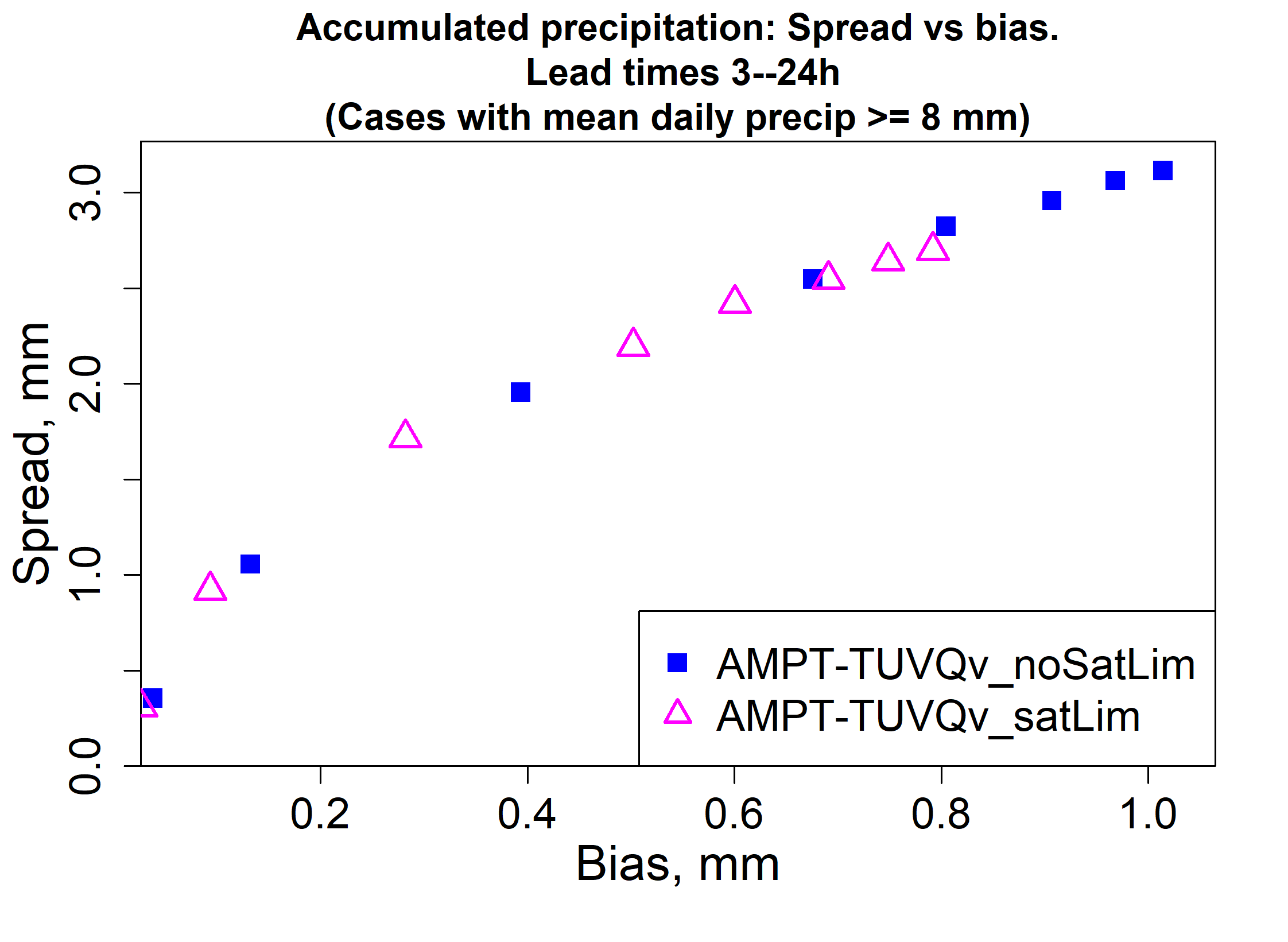}}
\end{center}
 \caption{Impact of the  supersaturation limiter on spread and bias of 
accumulated precipitation forecasts for lead times from 3 to 24 h. 
Averaging over cases with the mean (over the model domain) daily
accumulated precipitation  $\ge 1$ mm ({\em left}) and  $\ge 8$ mm ({\em right}).}
\label{Fig_satLim}
\end{figure}

First, we found that the impact on the near-surface temperature was 
rather small, so we focused on the impact on precipitation.
Figure \ref{Fig_satLim} shows the bias-spread scatterplots for two configurations of AMPT:
without and with the supersaturation limiter.
One can see that the bias-to-spread ratio was almost the same for the two configurations
whilst the spread was larger without  the supersaturation limiter
(a qualitatively similar result was obtained in the setting without humidity AMPT perturbations, not shown).
Therefore, we decided that the supersaturation limiter is not worth being
activated in AMPT.
So, for the experiments described in the rest of the article,  the supersaturation limiter
was off in both SPPT and AMPT.

\subsection {Roles of humidity and temperature perturbations}

Here we compare three configurations of AMPT and the basic configuration of SPPT in terms of their impacts
on the bias and the spread of the ensemble. In the \verb"AMPT-TUVQv" configuration,
$T,u,v,q_{\rm v}$ fields were perturbed. In the \verb"AMPT-TUV" configuration, only
$T,u,v$ fields were perturbed (that is, without humidity model perturbations). 
In the \verb"AMPT-UV" configuration, only
$u,v$  fields were perturbed. 
With  \verb"AMPT-UV", we found that the spread in the ensemble
was somewhat too small.
This led us to increase the amplitude multiplier $\kappa$ in this configuration from its
default value of 0.75 (see section \ref{sec_expm}) to 1. 

\begin{figure}[h]
\begin{center}
 \scalebox{0.44}{ \includegraphics{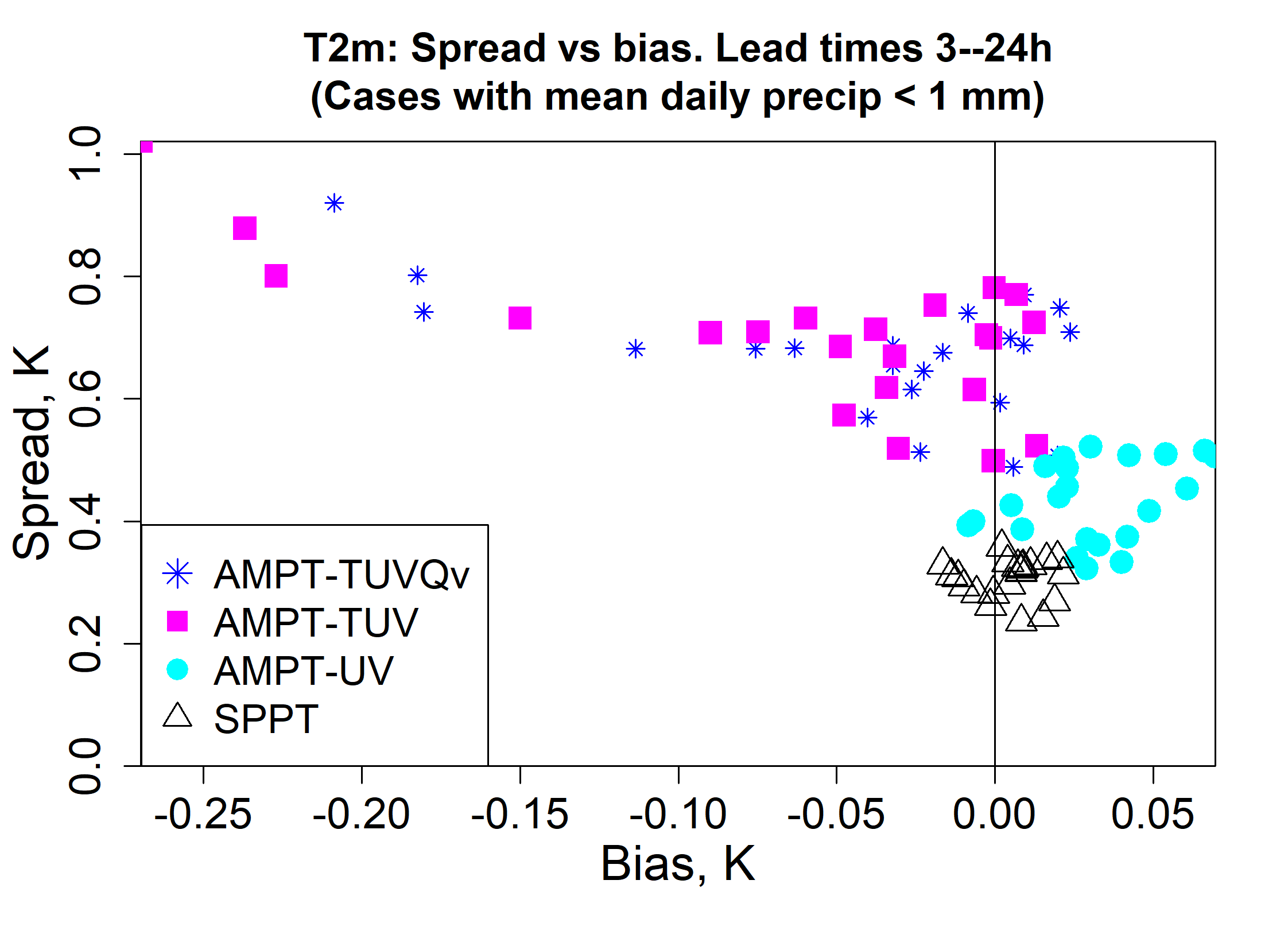}}
 \scalebox{0.44}{ \includegraphics{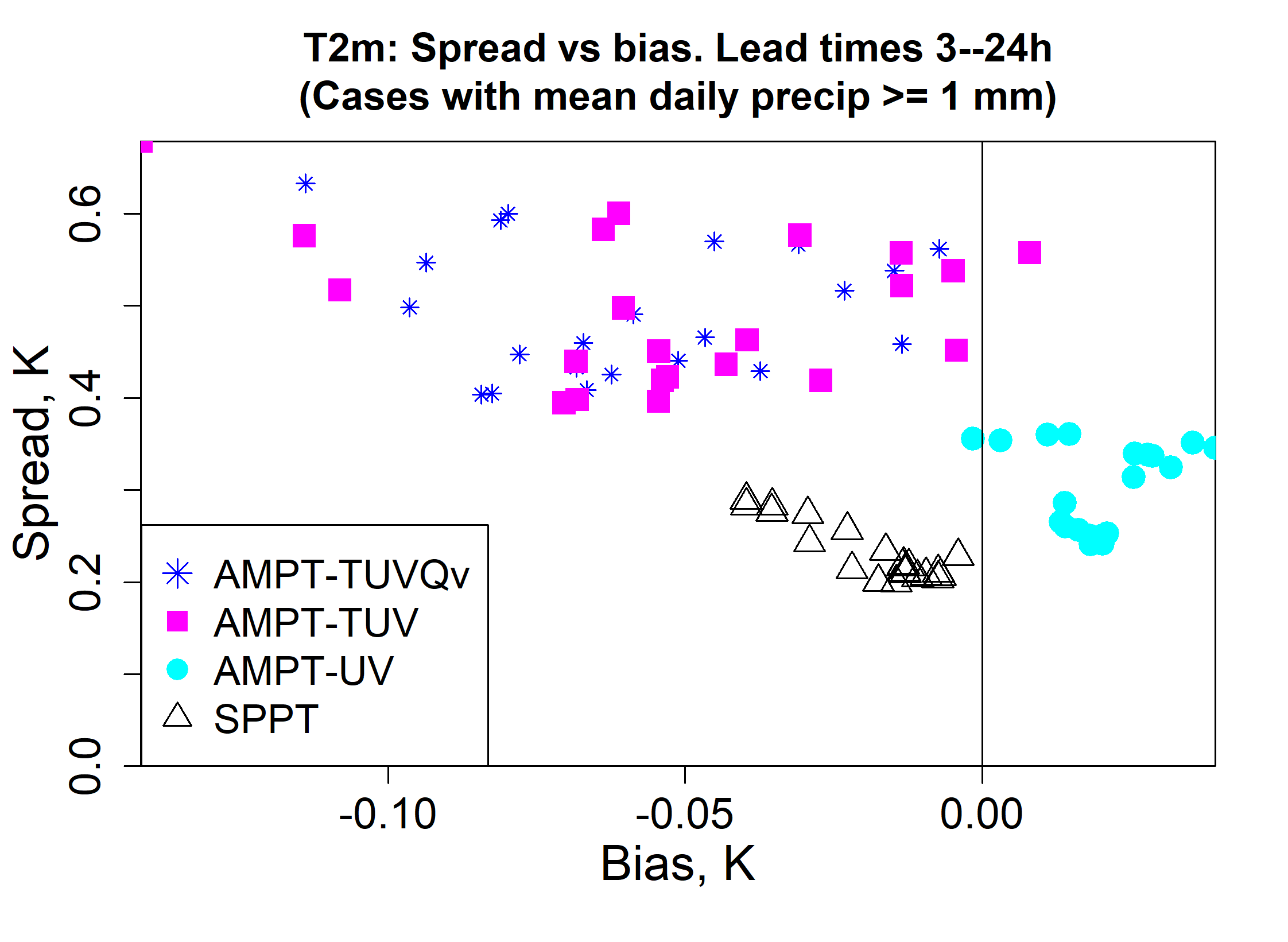}}
 \scalebox{0.44}{ \includegraphics{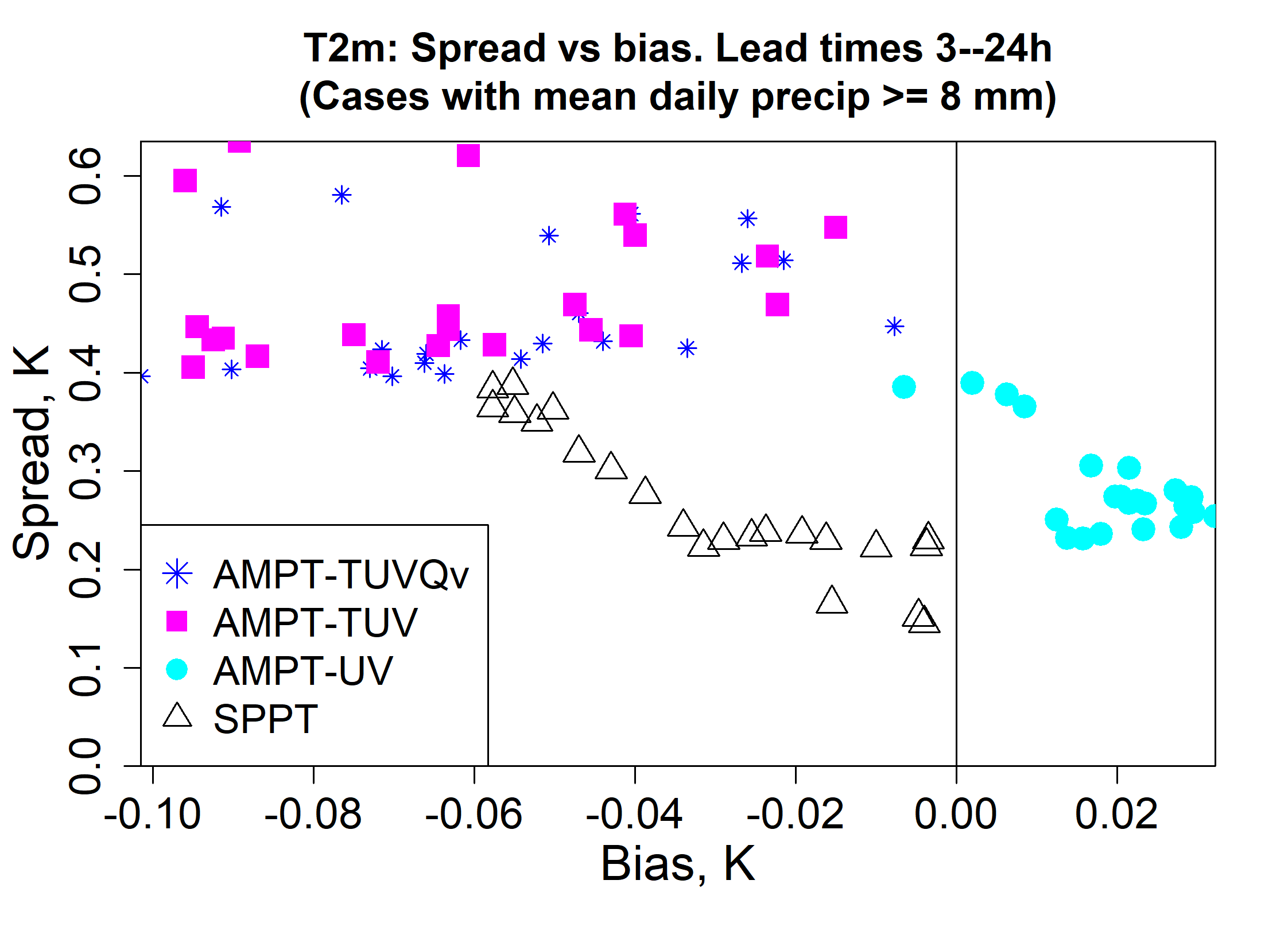}}
\end{center}
 \caption{Spread \vs bias of $T_{\rm 2m}$ forecasts for lead times from 3 to 24 h. 
Averaging over cases with the mean (over the model domain) daily
accumulated precipitation $<1$ mm ({\em top, left}), $\ge 1$ mm ({\em top, right}), and $\ge 8$ mm ({\em bottom}).}
\label{Fig_sprBias_T2m}
\end{figure}

Figure \ref{Fig_sprBias_T2m} shows bias-spread scatterplots for near-surface temperature.
One can see that perturbing humidity on top of $T,u,v$ did not change much neither the bias nor the spread
(\verb"AMPT-TUV" and \verb"AMPT-TUVQv" performed similarly in these experiments).
The other pair of schemes, \verb"SPPT" and \verb"AMPT-UV", had 
significantly lower (in modulus) bias and spread. 
In dry conditions (the top left plot), \verb"SPPT" had a smaller  spread and 
bias-to-spread ratio than \verb"AMPT-UV".
In wet conditions  (the top right plot),
\verb"SPPT" and \verb"AMPT-UV" performed similarly in terms of both spread and the bias-to-spread ratio,
but the signs of the biases were different: negative for SPPT and positive for AMPT.
In very wet conditions (the bottom plot), it was
\verb"AMPT-UV" that had the smallest   bias-to-spread ratio while having nearly the same spread as \verb"SPPT".

\begin{figure}[h]
\begin{center}
 \scalebox{0.44}{ \includegraphics{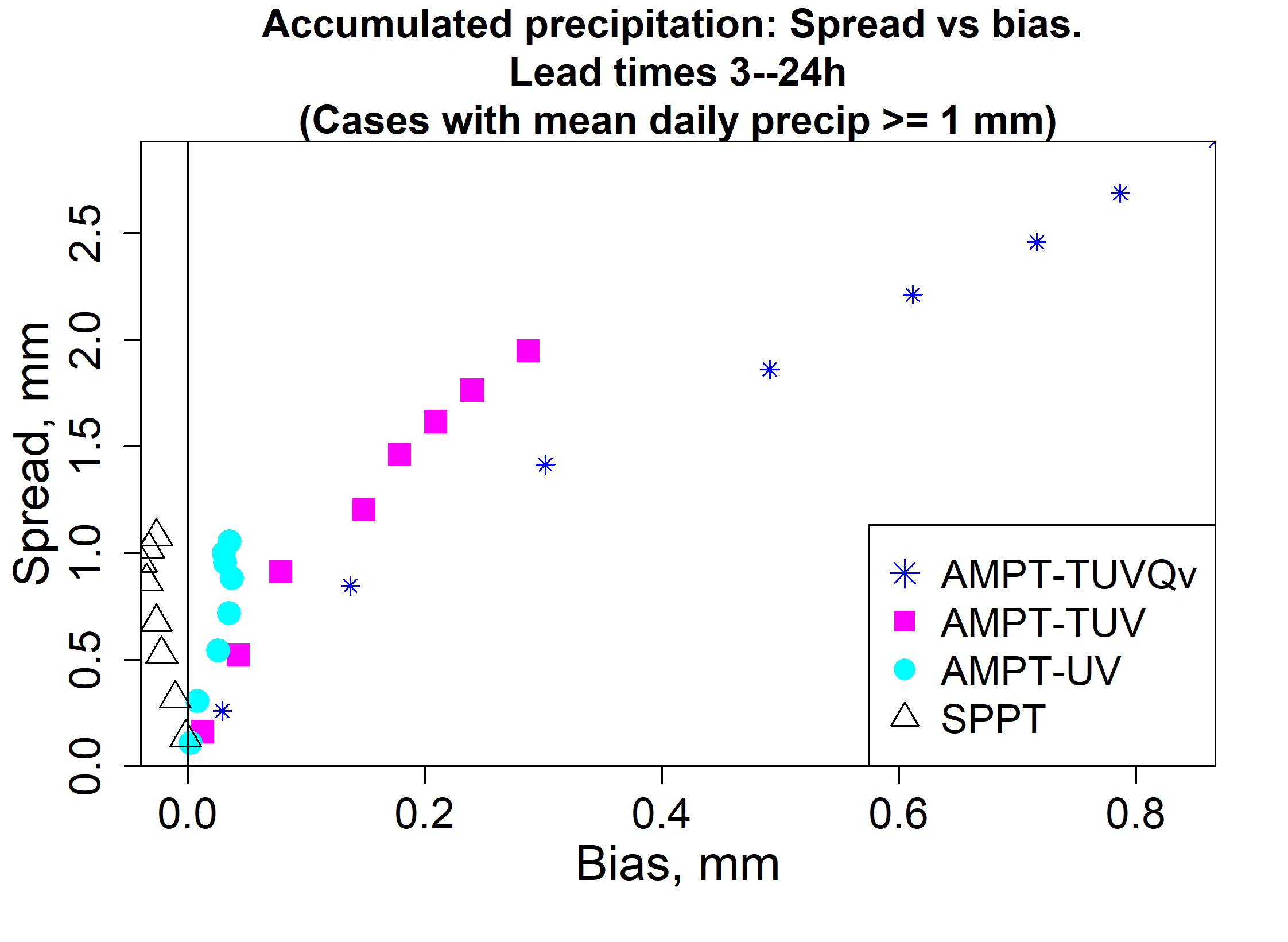}}
 \scalebox{0.44}{ \includegraphics{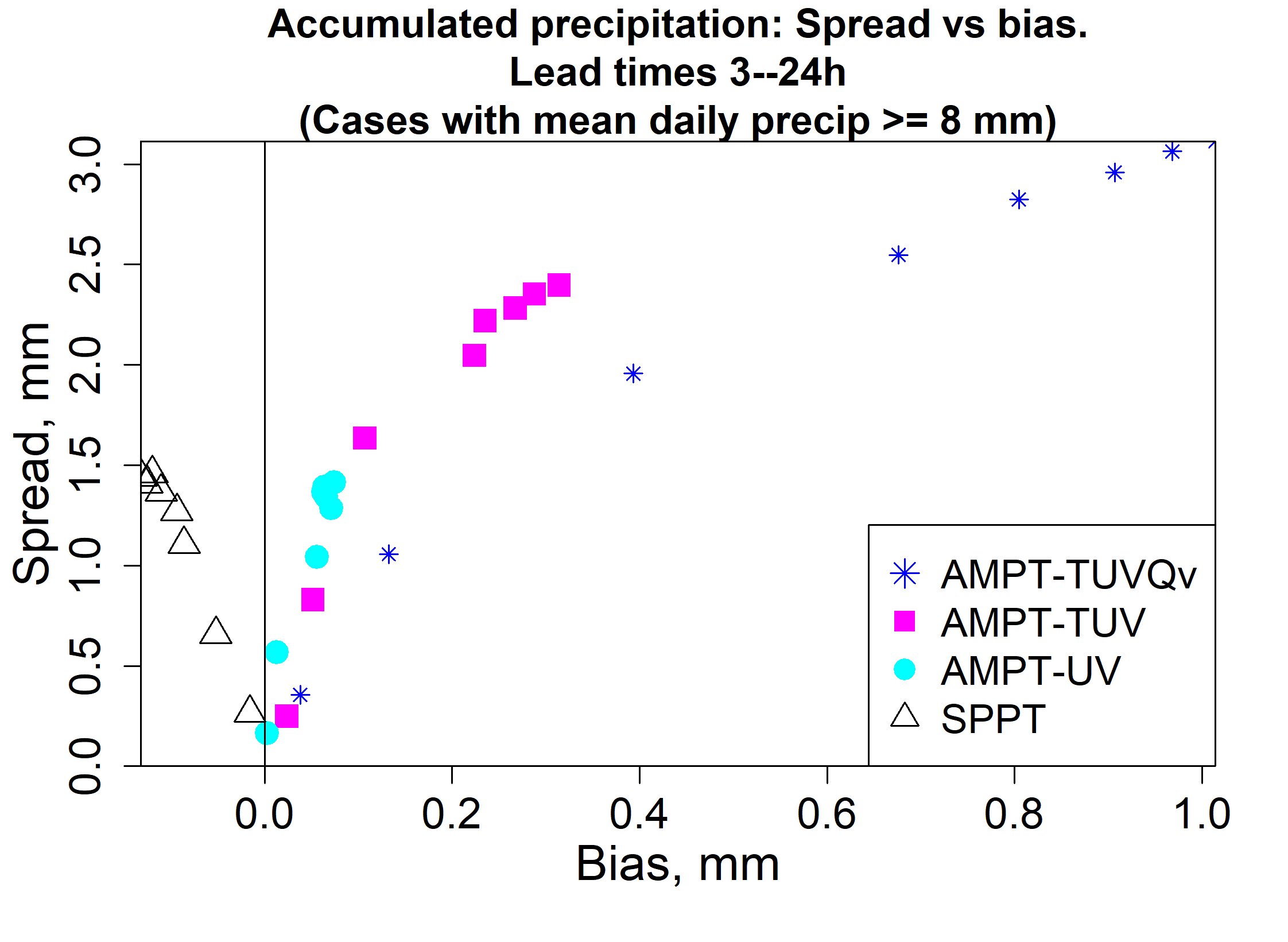}}
\end{center}
 \caption{Spread \vs bias of accumulated precipitation forecasts for lead times from 3 to 24 h. 
Averaging over cases with the mean (over the model domain) daily
accumulated precipitation $\ge 1$ mm (left) and $\ge 8$ mm (right).}
\label{Fig_sprBias_precip}
\end{figure}

Figure \ref{Fig_sprBias_precip} shows bias-spread scatterplots for accumulated precipitation.
In both  wet  (the left plot) and very wet (the right plot) conditions, the differences between
the four schemes were more systematic than in Fig. \ref{Fig_sprBias_T2m}.
Like in Fig. \ref{Fig_sprBias_T2m}, the magnitudes of precipitation
bias and spread of  \verb"SPPT" and \verb"AMPT-UV" were quite similar.
Again, like for $T_{\rm 2m}$, \verb"AMPT-TUV" and \verb"AMPT-TUVQv"  had significantly greater
(than \verb"SPPT" and \verb"AMPT-UV") bias and spread, 
but unlike Fig. \ref{Fig_sprBias_T2m}, the bias-to-spread  ratio for 
\verb"AMPT-TUV" was significantly lower than for \verb"AMPT-TUVQv".

It is also worth noting that the precipitation biases in Fig. \ref{Fig_sprBias_precip}
were negative for SPPT and positive for AMPT. Investigating mechanisms that lead a complex 
nonlinear model to generate biases of different signs under  different  zero-mean stochastic model perturbations
would be interesting but is beyond the scope of this research.

We summarize findings from Figs.  \ref{Fig_sprBias_T2m} and \ref{Fig_sprBias_precip} as follows.
\begin{enumerate}
\item
The magnitudes of forecast bias and  spread due to  \verb"SPPT" were similar to those due to 
\verb"AMPT-UV" (\ie AMPT without temperature and humidity perturbations).

\item
The addition of temperature perturbations in AMPT (in  \verb"AMPT-TUV")
led to significantly greater spread but somewhat higher (\ie worse) bias-to-spread ratio.

\item
Perturbing humidity on top of $T,u,v$ (in  \verb"AMPT-TUVQv") 
led to a significant growth in spread for precipitation but at the expense of 
a significant degradation in the bias-to-spread ratio.

\end{enumerate}

Thus, without temperature perturbations in AMPT (\ie perturbing only $u,v$), 
we could not outperform SPPT in terms of spread.
However, with humidity perturbations in AMPT  (\ie perturbing $T,u,v,q_{\rm v}$), 
forecast biases for precipitation were too large.
Therefore, in the ensemble prediction experiments described in the next subsection,
AMPT temperature (and winds) perturbations were switched on whereas 
humidity perturbations were switched off.

\section{Testing AMPT in the ensemble prediction system}
\label{sec_expm_eps}

\subsection{Setup}
\label{sec_setup_EPS}

The general experimental setup described in section \ref{sec_expm} applies to the experiments presented in this section
with the caveat that there was no unperturbed member in the ensemble here. The 
results were verified  against near-surface observations (about 40 stations) using 
the    VERification System Unified Survey  (VERSUS) developed within the COSMO consortium \citep{Gofa}. 

The list of experiments and their basic features are presented in Table \ref{Tab_list_expm}. 
In \verb"AMPT-SOIL", initial soil perturbations generated following section \ref {sec_soil_ini}
were added to the respective fields of the initial-conditions ensemble members.

\begin{table}[h]
\caption{List of ensemble prediction experiments}

\begin{center}
\begin{tabular}{lll}

\hline
Experiment & Model perturbations \\
\hline
\verb"NOPERT"        &   None  \\
\verb"SPPT"          &   Atmospheric SPPT perturbations  \\
\verb"AMPT-NOSOIL"   &   Atmospheric AMPT perturbations  \\
\verb"AMPT-SOIL "    &   Atmospheric and soil AMPT perturbations \\
\hline
\end{tabular}
\end{center}
\label{Tab_list_expm}
\end{table}

Besides the four model perturbation schemes listed in Table \ref{Tab_list_expm},
we also tested a hybrid of AMPT and SPPT. 
We expected that such a hybrid might lead to an improvement because SPPT and
AMPT perturbations explore different volumes in phase space. 
Indeed, the direction of the SPPT perturbation vector 
(comprised of all model variables at a grid point or a grid column)
exactly coincides with the direction
of the physical tendency vector, Eq.(\ref{d_SPPT}).
In contrast, the AMPT perturbation vector
can have any direction because $\xi_i$ are uncorrelated, see  Eq.(\ref{d_AMPT}) and
section \ref{sec_multivar}, so it
has nothing to do with the direction of the model's physical tendency vector at all.
Besides, the magnitude of the SPPT perturbation is proportional to the magnitude of the 
local physical tendency, which is never the case with  AMPT.
Somewhat surprisingly, we could not see any benefit from the hybridization
of the two schemes (not shown), so we abandoned the hybrid.
As suggested by an anonymous reviewer, this outcome might be caused by 
the smaller magnitude of SPPT perturbations,
which therefore could not make a difference.

During the two-month trial, there were two cases (out of 59) in which one of the forecasts exhibited catastrophic instability. 
On March 15, three out of the four  tested configurations  `exploded': \verb"NOPERT" 
(in which, we recall, there were no model perturbations, only initial and lateral-boundary 
conditions were perturbed), 
\verb"SPPT", and \verb"AMPT-NOSOIL",
whereas the most successful AMPT configuration, \verb"AMPT-SOIL", performed normally.
On March 20, both AMPT configurations failed whereas \verb"NOPERT" and \verb"SPPT" were OK.
These two cases were excluded from the forecast performance statistics.
So, our experiments showed that numerical instability can happen with AMPT, but its frequency
is nearly as low as with SPPT.  A longer test period is needed to make a more precise statement.

In this section, we show  verification scores for $T_{\rm 2m}$ forecasts averaged over the whole two-month period. 
Similar results were obtained for each of the two months separately 
(not shown). 
For near-surface wind speed  $V_{\rm 10m}$, 
verification results were, mostly, similar to those for $T_{\rm 2m}$, except that
soil perturbations had a significantly smaller impact on the wind speed than on temperature. This
could be expected because perturbations of
soil temperature and soil moisture directly impact surface fluxes of sensible and latent heat
but not the flux of momentum.
Due to the similarity of verification scores for $V_{\rm 10m}$  compared to $T_{\rm 2m}$,
we only display one typical plot for $V_{\rm 10m}$ below.
We also present a comparison of AMPT and SPPT for upper-level fields.

With precipitation, the only statistically significant effect of AMPT
was an increase in ensemble spread.
This can be seen in the model-perturbations-only experiments reported above in section \ref{sec_hum_bias} 
(see Fig. \ref{Fig_sprBias_precip})
and this was observed in the experiments described in this section (not shown).
Other verification scores for precipitation forecasts were ambiguous because, on the one hand,
the time period  was rather dry and on the other hand, the observation network
was too scarce to detect relatively rare and localized precipitation events (not shown).

In contrast to  section \ref{sec_hum_bias}, where the results were obtained by averaging over 
large samples (the whole spatial grid and two months of data), 
the ensemble forecasts examined in this section were verified against a relatively scarce
observation network. This caused a larger statistical uncertainty and therefore required 
statistical significance testing, see Appendix \ref{App_signif} for methodological details.
The default settings of the statistical significance calculations were as follows.
We examined statistical significance of improvements due to 
 \verb"AMPT-NOSOIL" relative to  \verb"SPPT" because both schemes 
involved no soil perturbations and thus were comparable. 
We averaged the scores over all lead times up to 48 h to reduce statistical uncertainty.
The significance level  was set at  $\alpha=0.05$.
The number of bootstrap samples was $10^5$.
Along with the $p$-value of the above bootstrap test, $p_{\rm bootstrap}$,
we also computed the $p$-value of the classical one-sided Student's t-test, $p_{\rm Student}$.

\subsection{Accuracy of ensemble mean forecasts and reliability of probabilistic forecasts}
\label{sec_reliab}

\begin{figure}[h]
\begin{center}
 \scalebox{0.58}{ \includegraphics{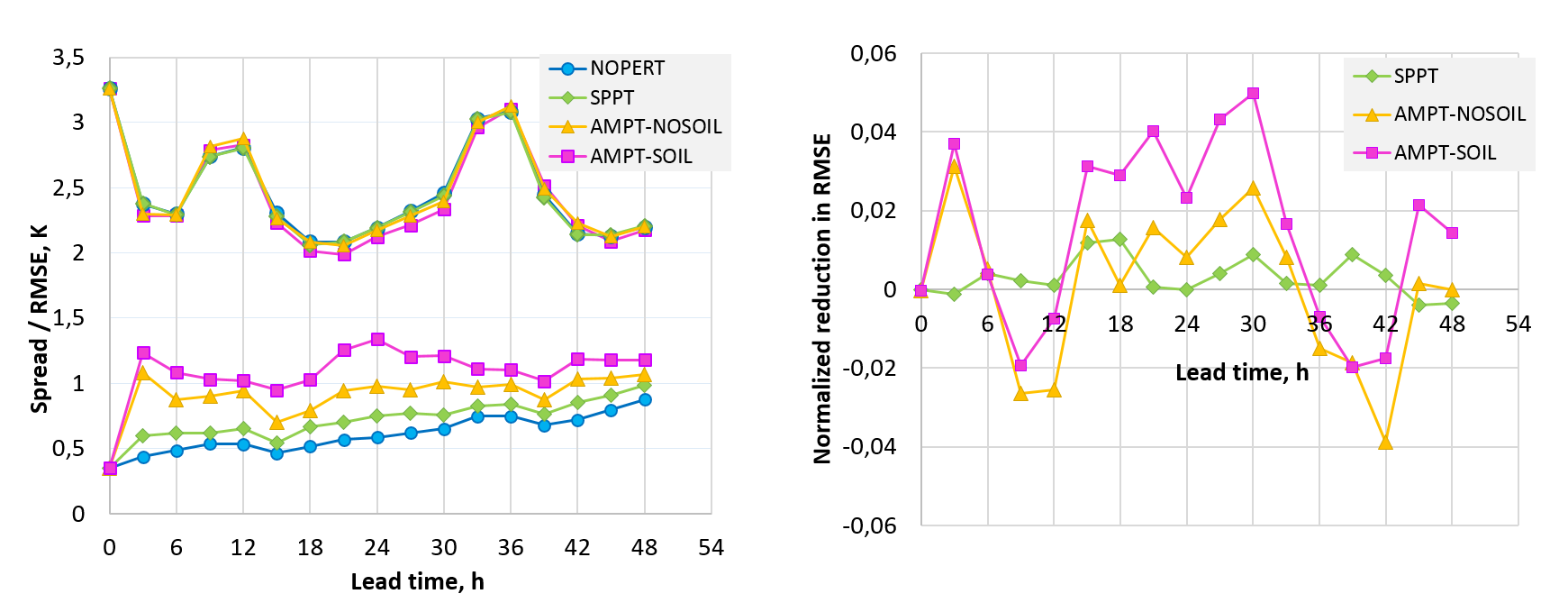}}
\end{center}
 \caption{$T_{\rm 2m}$. {\em Left}: RMSE of ensemble mean (the upper bunch of curves) and 
 ensemble spread (the lower bunch of curves).
{\em Right}:  The normalized reduction in  ensemble-mean RMSE, that is,
${\rm (RMSE_{NOPERT} - RMSE)/RMSE_{NOPERT}}$, the higher the better.
}
\label{Fig_RMSE}
\end{figure}

Figure \ref{Fig_RMSE}(left) shows  root-mean-square errors
(RMSE) of the $T_{\rm 2m}$  ensemble mean forecast (the upper bunch of curves). 
Each curve corresponds to a 
model perturbation scheme from the list in Table \ref{Tab_list_expm}.
One can see in Fig.  \ref{Fig_RMSE}(left) that, excluding the initial transient (spinup) 
period of some 3 h (likely, due to imbalances
in initial conditions), 
the  RMSE had a prominent diurnal cycle with a broad minimum 
at night and a narrower maximum shortly after midday.

To highlight the barely seen in Fig. \ref{Fig_RMSE}(left)
differences between the curves in the upper bunch,
Fig. \ref{Fig_RMSE}(right) displays the normalized reduction in the 
$T_{\rm 2m}$ ensemble-mean RMSE  with respect  to \verb"NOPERT",
that is,
${\rm (RMSE_{NOPERT} - RMSE)/RMSE_{NOPERT}}$.
SPPT perturbations led to a slight overall reduction in RMSE.
\verb"AMPT-NOSOIL"  perturbations gave rise to a more significant decrease in the $T_{\rm 2m}$ RMSE 
most of the time except  for the rather  short time period in the afternoon local time when the
ensemble-mean forecast deteriorated.
The effect of soil perturbations on the ensemble mean forecast looks positive: the 
RMSE reduction  is seen to be nearly uniformly higher in the experiment \verb"AMPT-SOIL" than in  \verb"AMPT-NOSOIL". 
However, all these differences were not statistically significant at the 0.05 level.
For near-surface wind speed, the results were similar, not shown.

The  lower bunch of curves in Fig. \ref{Fig_RMSE}(left)
shows the ensemble spread in $T_{\rm 2m}$ for the four schemes examined. 
One can see that the spread was somewhat too low  with both SPPT and AMPT,
though AMPT perturbations induced substantially greater spread than SPPT perturbations
(this conclusion was valid for the spread in  $V_{\rm 10m}$ as well, not shown).
The advantage of \verb"AMPT-NOSOIL" over  \verb"SPPT"  
in terms of spread was highly statistically significant: 
the $p$-values of both the bootstrap test and the Student's t-test were less than $0.001$.
For all 8 weeks and all lead times, the forecast spread with \verb"AMPT-NOSOIL" 
was substantially greater than the spread with  \verb"SPPT".
The advantage of \verb"AMPT-SOIL"  over  \verb"SPPT" is seen to be even bigger.

Note that in an ideal ensemble, its members are indistinguishable from the truth.
Both  members of such an ensemble and the truth can be viewed as independent draws from the same
probability distribution.
In particular, the expectation of any member equals  the expectation of the truth. 
The standard deviation  of any member at some point in space and time 
(let it be denoted by $\sigma$) is  the same as the respective
standard deviation  of the truth.
Then, the expected square of the ensemble spread 
(defined as the square root of the unbiased sample variance) equals $\sigma^2$ by construction.
At the same time, the expected square of the error in the ensemble mean 
(the mean-square error, MSE) equals $\frac{N+1}{N}\sigma^2$, where $N$ is 
the ensemble size \citep{Fortin}.
Being averaged in space and time, the expected square of the ensemble spread becomes equal
to $\overline{\sigma^2}$.
Correspondingly, the averaged in space and time MSE becomes $\frac{N+1}{N}\overline{\sigma^2}$.
Thus, it is clear that for large ensembles, the root-mean-square spread should be very close to RMSE. For an ensemble
of size $N=10$ used in this study, RMSE should, on average, exceed the   spread
by $\sqrt{\frac{N+1}{N}} -1 \approx 0.05$, which is of course much smaller than the differences
between the RMSE and the spread seen in Fig. \ref{Fig_RMSE}(left).

Another reason why forecast RMSE can be greater than ensemble spread is observation error.
In complex terrain, the dominant source of error can be {\em representativeness uncertainty},
which accounts for the fact that observations can poorly represent grid-cell
averaged fields provided by the model.
We did not account for the contribution of observation error to RMSE.

It is worth mentioning that the variations in the RMSE as functions of lead time 
seen in Fig. \ref{Fig_RMSE}(left) are not accompanied by corresponding variations in the spread.
The reason is that the diurnal-cycle variations in RMSE were largely  caused  
by systematic forecast errors (biases). Indeed,
verification results of COSMO model forecasts on different
domains presented in \citep{Rieger} show that the magnitude of diurnal variations in the forecast bias 
is between 1K and 2K, with a broad maximum at night and a narrow minimum in the afternoon. 
{\em Stochastic} zero-mean model perturbations, cannot, in general,  represent model biases,
implying that the forecast spread cannot reflect the contribution  of the bias to the RMSE.
Either the forecasts are to be debiased or a multi-physics/multi-model ensemble 
(in which different members have different biases)
is to be used, see, \eg \citet{Berner2015}.

\begin{figure}[h]
\begin{center}
 \scalebox{0.44}{ \includegraphics{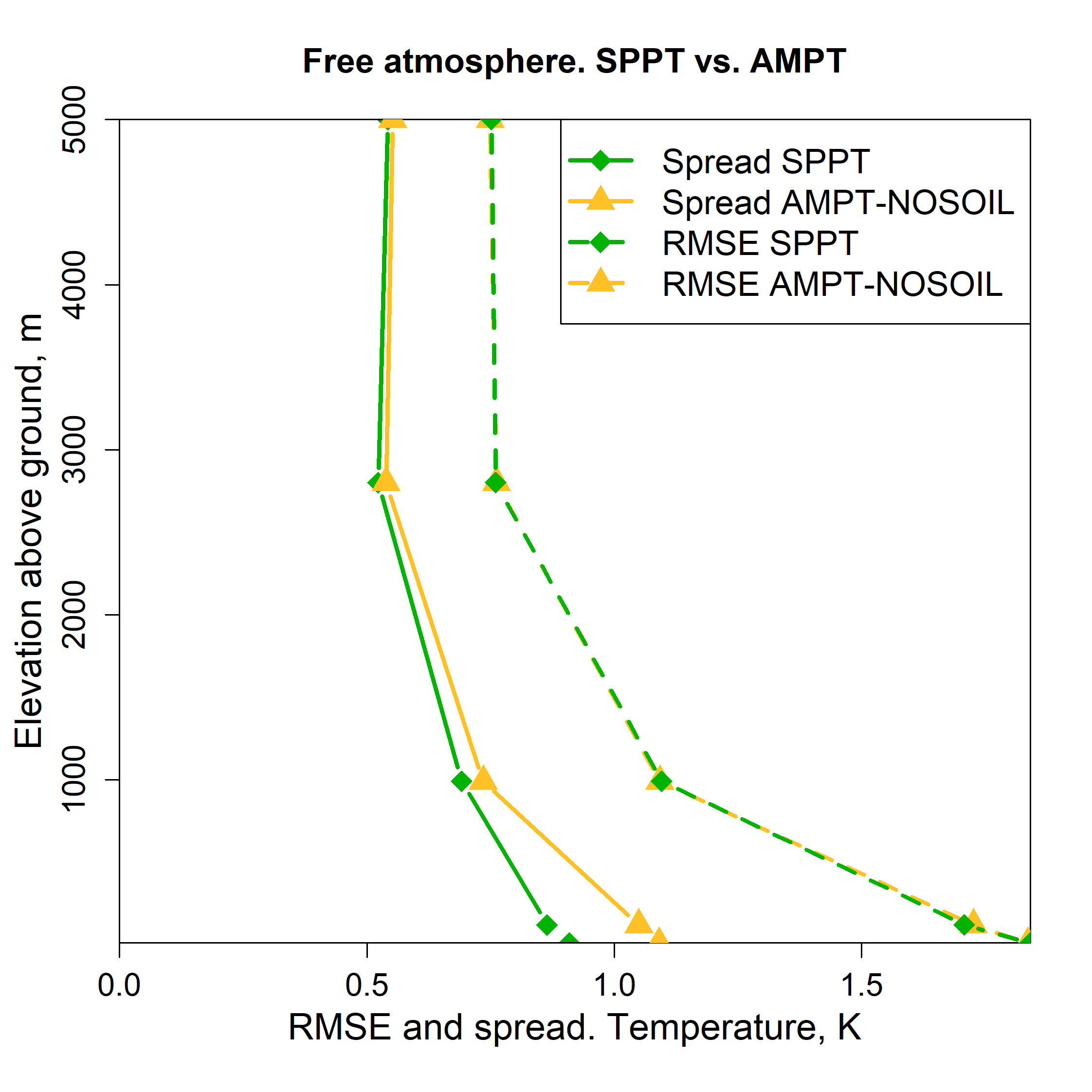}}
 \scalebox{0.44}{ \includegraphics{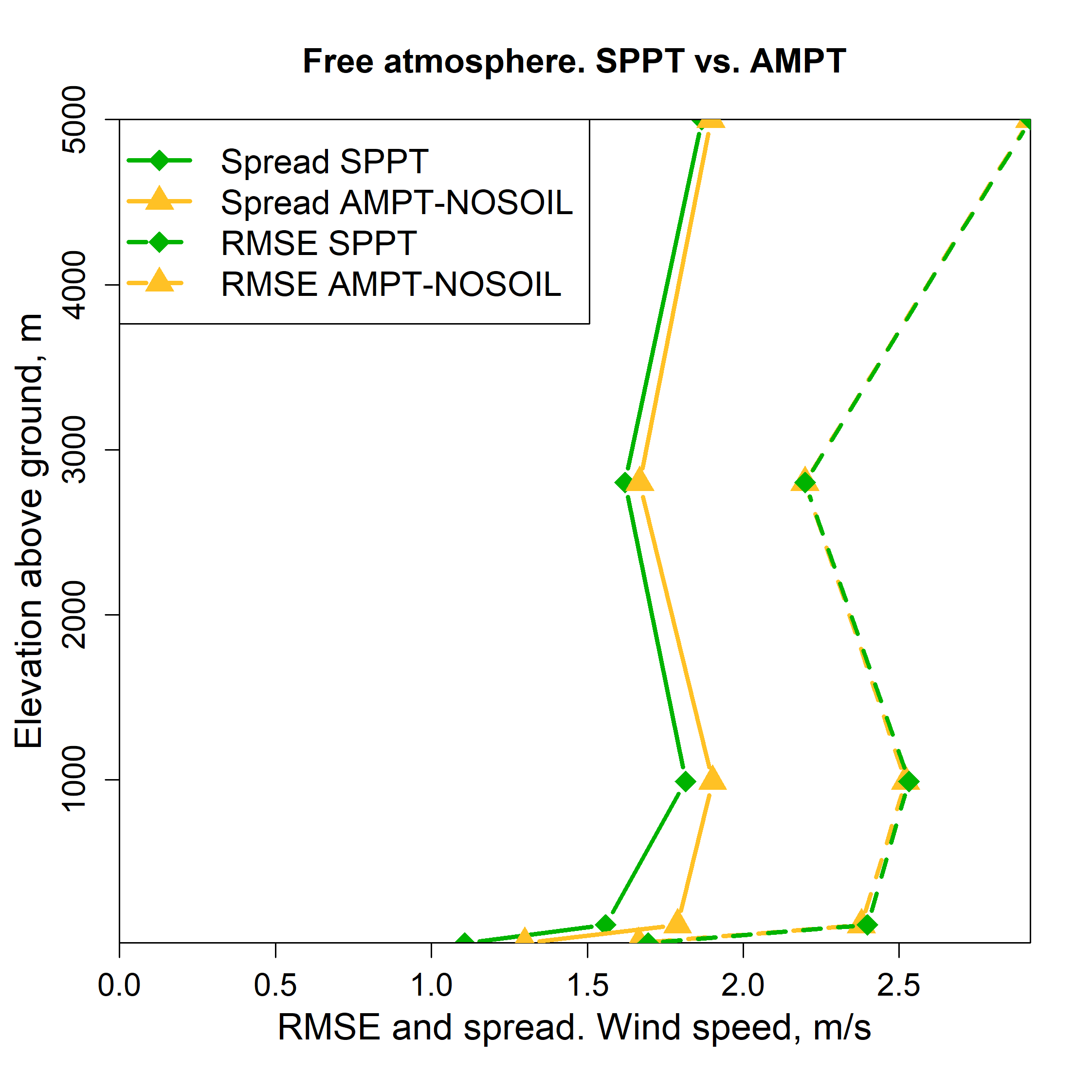}}
\end{center}
 \caption{RMSE and spread for AMPT-NOSOIL vs. SPPT in free atmosphere.\\
   {\em Left}: Temperature.
{\em Right}:  Wind speed.}
\label{Fig_upper}
\end{figure}

Higher in the atmosphere, the effects of AMPT  were similar 
to what we found for the near-surface fields.
Figure \ref{Fig_upper} shows the RMSE and spread for upper-level temperature and wind speed
(averaged over four lead times: 12, 24, 36, and 48 h).
RMSE was computed here with respect to the ECMWF analyses interpolated to the model grid.
It is seen that AMPT produced
a  higher spread and nearly the same RMSE as SPPT.
The superiority of AMPT over SPPT in terms of spread was statistically significant 
for both fields at all levels 
shown in Fig.\ref{Fig_upper}, with $p<0.001$ for both the bootstrap test and Student's t-test. 
It is interesting to note that for temperature, the RMSE and spread were maximal near the ground
(where turbulence and small-scale disturbances due to orography are the strongest). 
For wind speed, on the contrary, both the RMSE and spread were the smallest at the lowest model levels.
Their downward decrease was most pronounced in the surface layer
(perhaps, because the wind speed itself is relatively low near the ground due to turbulent friction).

If the spread is systematically and significantly different (say, lower, as in our experiments) 
than  RMSE, 
then the conditional cumulative distribution function of the verification (truth, observation) $x_{\rm v}$ 
given the ensemble cumulative distribution function $F_{\rm e}(x)$ of the model variable $x$,
\ie $P(x_{\rm v} {\le} x \mid F_{\rm e})$ (where $P$ stands for probability), is
systematically different from $F_{\rm e}(x)$. 
This kind of inconsistency is known as lack of {\em reliability}
of the probabilistic forecast \citep[][]{Toth}.
In these terms, Figs. \ref{Fig_RMSE}(left) and \ref{Fig_upper} demonstrate that
AMPT perturbations are capable of significantly improving reliability of  ensemble-based
probabilistic forecasts compared to SPPT perturbations.

\begin{figure}[h]
\begin{center}
 \scalebox{0.75}{ \includegraphics{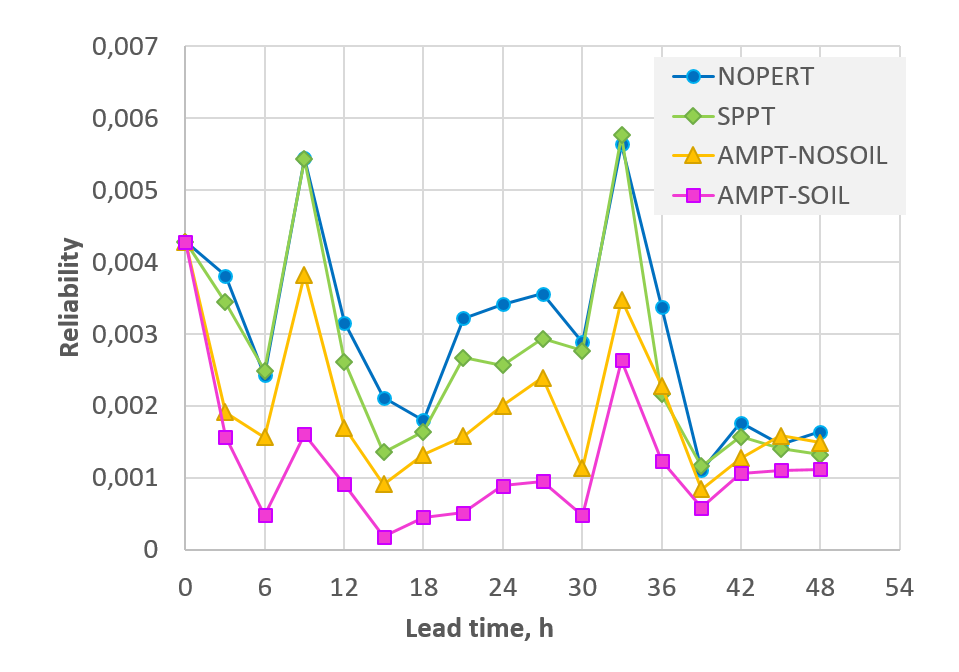}}
\end{center}
 \caption{The reliability component of the Brier score for the event $T_{\rm 2m}>0^\circ C$. The lower the better.}
\label{Fig_reliabBrier}
\end{figure}

Figure \ref{Fig_reliabBrier} presents a more direct illustration of improvements in reliability brought about by
 \verb"AMPT-NOSOIL" and \verb"AMPT-SOIL"
with respect  to \verb"NOPERT" and \verb"SPPT". 
It shows the reliability component of the Brier score\footnote{
The reliability component of the Brier score is
a weighted mean-square difference between the predicted event probabilities $p^{\rm f}_k$ 
(where $k$ labels  the discrete values of the forecast probability obtainable from a 
finite-size ensemble) and the respective 
observed frequencies of the event for all cases when the  probability $p^{\rm f}_k$ was forecast.}
 for the (most populated) event
 $T_{\rm 2m}>0^\circ C$.
The improvement of \verb"AMPT-NOSOIL" with respect to \verb"SPPT"
was statistically significant, $p_{\rm bootstrap}=0.05$ (though $p_{\rm Student} = 0.07$ did not reach
the significance level of 0.05).

\subsection{Resolution and discrimination of probabilistic forecasts}
\label{sec_res_discr}

A {\em reliable} uncalibrated ensemble-based probabilistic forecast tells 
the user that if the ensemble variance (\ie the spread squared)
at some point in space and time equals some number $d$, then
the expected MSE (RMSE squared) 
also approximately equals $d$.
This is an important property expected from an ensemble. However, 
reliability alone does not fully characterize the quality of a probabilistic forecast.
Indeed, the constant spread equal to the `climatological' RMSE would imply a perfectly reliable forecast,
which is, however, not much useful.
The ensemble needs to generate sufficiently variable spread. 
If it does and if variations in the spread correspond to variations in the RMSE
(the property known as {\em resolution}), then the ensemble can provide
relevant information about the spatially and temporally variable RMSE.

\begin{figure}[h]
\begin{center}
 \scalebox{0.7}{ \includegraphics{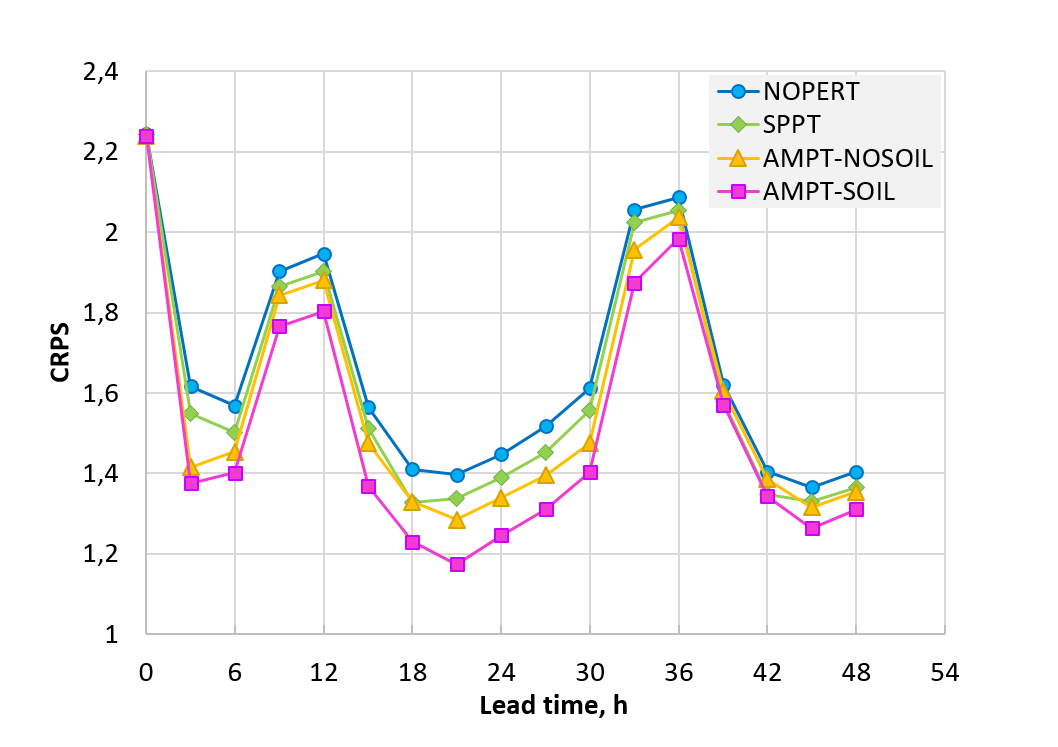}}
\end{center}
 \caption{CRPS for $T_{\rm 2m}$. The lower the better. Note that the y-axis does  not start at  0.}
\label{Fig_CRPS}
\end{figure}

One useful measure of resolution  combined with reliability is the continuous ranked probability score (CRPS).
Figure \ref{Fig_CRPS} displays CRPS for $T_{\rm 2m}$. 
One can see that both SPPT and AMPT perturbations 
did improve CRPS (compared with \verb"NOPERT") 
while AMPT led to substantially greater improvements, 
especially with additional soil perturbations. 
The improvements of  \verb"AMPT-NOSOIL"  compared with  \verb"SPPT"
were statistically significant. The $p$-value of the 
bootstrap test computed following Appendix \ref{App_signif} was $p_{\rm bootstrap}=0.03$.
The $p$-value of the Student's t-test was $p_{\rm Student} = 0.02$.
\verb"AMPT-SOIL"
performed {\em a fortiori} statistically significantly better than \verb"AMPT-NOSOIL".

\begin{figure}[h]
\begin{center}
 \scalebox{0.7}{ \includegraphics{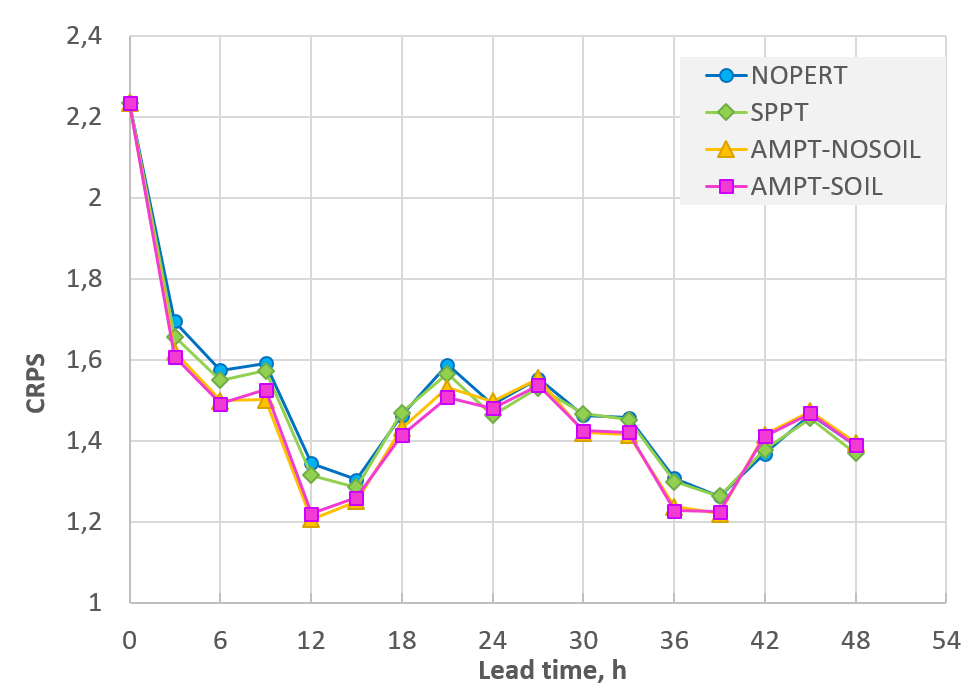}}
\end{center}
 \caption{CRPS for  $V_{\rm 10m}$. The lower the better. Note that the y-axis does  not start at  0.}
\label{Fig_CRPS_wind}
\end{figure}

Figure \ref{Fig_CRPS_wind} displays CRPS for near-surface wind speed  $V_{\rm 10m}$.
The impact of AMPT on the forecast performance in terms of  $V_{\rm 10m}$ was weaker 
than on  $T_{\rm 2m}$ but more stable, so the statistical
significance of the advantage of \verb"AMPT-NOSOIL" \vs \verb"SPPT"
was stronger: $p_{\rm bootstrap}<0.001$ and $p_{\rm Student} = 0.004$. 
Soil perturbations contributed little to the forecast CRPS. We mentioned a reason for that in section \ref{sec_setup_EPS}.
Why the impact of AMPT 
on wind was weaker (but more stable) than the impact on temperature in the settings without soil perturbations
remains unclear.

\begin{figure}[h]
\begin{center}
 \scalebox{0.7}{ \includegraphics{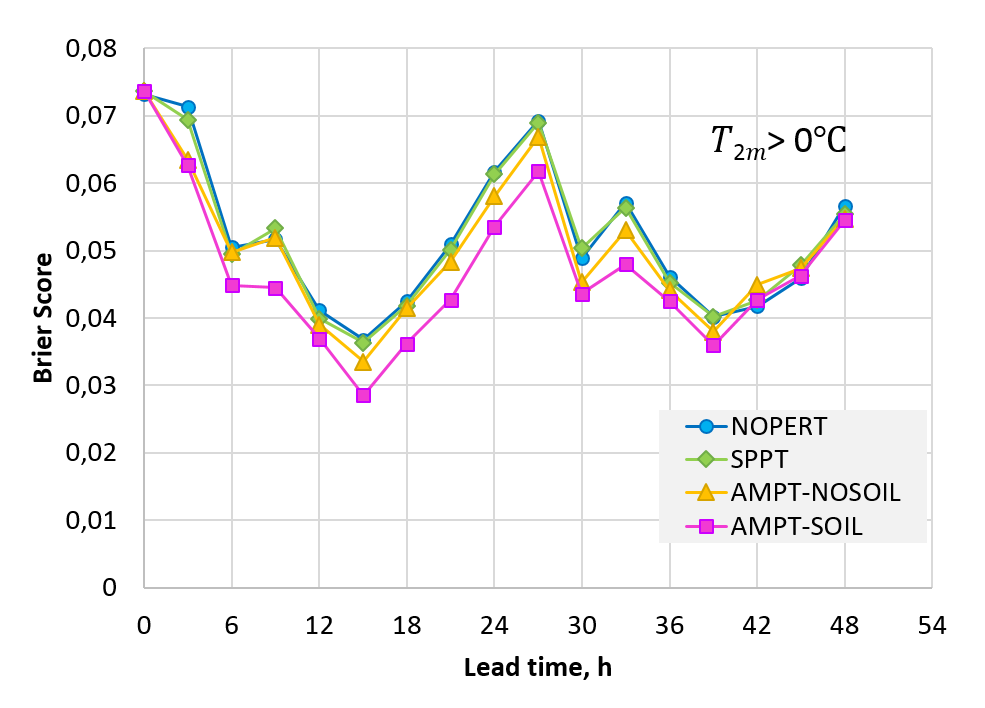}}
\end{center}
 \caption{Brier score for the event $T_{\rm 2m}>0^\circ C$.
  The lower the better.}
\label{Fig_Brier}
\end{figure}

Another popular measure of reliability and resolution is the Brier score, 
which, in contrast to CRPS, is computed for specific {\em events}. 
If the event is defined as $x<\theta$ (or $x > \theta$), where $x$ is the continuous model
variable in question (say, temperature) and $\theta$ is a threshold, then
CRPS is the integral of the Brier score over all possible $\theta$
\citep{Hersbach}. 
Therefore, Fig. \ref{Fig_CRPS} implies that the Brier score 
integrated over all temperature thresholds  is better with AMPT than with SPPT.
To find out what happens for specific distinct thresholds, we selected several
meteorologically relevant events ($T_{\rm 2m}<-5^\circ C$, $T_{\rm 2m}>0^\circ C$, $T_{\rm 2m}>5^\circ C$)  
and verified the ensemble forecasts
using the Brier score. AMPT did outperform SPPT for all these events, we show the results
for the most populated event $T_{\rm 2m}>0^\circ C$ in Fig. \ref{Fig_Brier}
(for near-surface wind speed, the results were similar, not shown).
The improvements of  \verb"AMPT-NOSOIL" (and \verb"AMPT-SOIL" as well) compared with  \verb"SPPT"
were again statistically significant: $p_{\rm bootstrap}=0.03$ and $p_{\rm Student} = 0.01$.
Similar results were obtained for near-surface wind speed, not shown.

We also examined the resolution component of the Brier Score.
With the reliability component substantially improved, see Fig. \ref{Fig_reliabBrier},  the 
resolution component was barely changed or slightly reduced (\ie degraded, not shown). 
As noted by \citet{Talagrand}, an increase in spread may lead to a degradation
in resolution because a larger spread is more akin to the climatological spread, which 
yields no resolution.
So, we state that AMPT substantially improved reliability without significantly degrading resolution.

As  discussed above, the ability of a probabilistic forecast 
to predict the uncertainty in the forecast  can be
measured by reliability and resolution  
through the conditional distribution of the verification $x_{\rm v}$ 
given the ensemble cumulative distribution function $F_{\rm e}(x)$, \ie $p(x_{\rm v} \mid F_{\rm e})$
(where $p$ stands for probability density).
The reciprocal and complementary view on the accuracy of a probabilistic forecast 
is  through $p(F_{\rm e} \mid x_{\rm v})$, \ie conditional on verification.
The capability of a probabilistic forecast to concentrate the probability mass close to the observed $x_{\rm v}$
(thus, discriminating between different outcomes $x_{\rm v}$)
is called {\em discrimination} \citep[e.g.,][]{Wilks}.

\begin{figure}[h]
\begin{center}
   \scalebox{0.7}{ \includegraphics{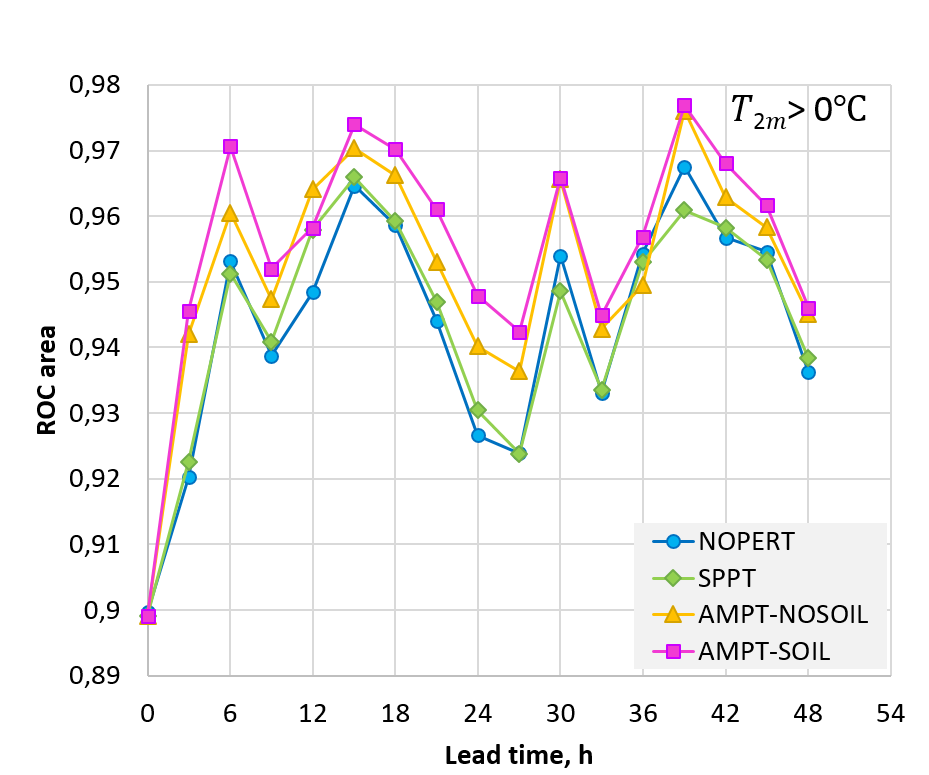}}
\end{center}
 \caption{ROC  area for the event $T_{\rm 2m}>0^\circ C$. 
  The higher the better. Note that the y-axis does  not start at  0.}
\label{Fig_ROCA}
\end{figure}

Discrimination is commonly assessed using the Relative Operating Characteristic (ROC) defined for 
a specific meteorological event and, more succinctly,  using the area under the ROC curve
(ROC area).
 Figure \ref{Fig_ROCA} shows the ROC area for the event $T_{\rm 2m}>0^\circ C$
 (similar results were obtained for the two other events, 
$T_{\rm 2m}<-5^\circ C$ and $T_{\rm 2m}>5^\circ C$, not shown). 
 From this figure, we see that AMPT did outperform SPPT.
Our statistical significance tests showed that the advantage of  \verb"AMPT-NOSOIL"  over \verb"SPPT"
in terms of ROC area was statistically significant: $p_{\rm bootstrap}=p_{\rm Student} = 0.01$.
As \verb"AMPT-SOIL" is seen in Fig.  \ref{Fig_ROCA} to be almost uniformly better than \verb"AMPT-NOSOIL", 
we conclude that \verb"AMPT-SOIL" was statistically significantly more skillful
for near-surface temperature than SPPT, too.
For near-surface wind speed, the results were mixed (not shown).

A few remarks regarding Fig.  \ref{Fig_ROCA} are in order.
First, the overall level of the ROC area score was quite high
(its perfect value is 1 whereas a value less than 0.5 indicates no skill) so 
the probabilistic forecasts in question were quite   skillful. 
Second, as with many other plots above, the ensemble forecasts were the
worst at zero lead time. This was, most likely, caused by the poor quality of the initial ensemble.
Finally, we note that the ROC area is known to be insensitive to the degree of reliability and cannot 
be improved by calibration of a probabilistic forecast \citep[e.g.,][]{Wilks}.
Therefore,
the superiority of \verb"AMPT-SOIL" over  \verb"SPPT"  and \verb"NOPERT"
in terms of ROC area implies that 
AMPT did not just inflate spread (which could be done by calibration),
it can yield `good spread' in the sense of \citet{EckelMass}, meaning greater spread when and where it is relevant.

\subsection{Summary of ensemble prediction results}

For near-surface temperature forecasts,
atmospheric AMPT  perturbations gave rise to statistically significant improvements (compared to SPPT) 
in the performance of the ensemble prediction system --- in terms of spread-skill relationship, 
CRPS, Brier score, and ROC area. 
In terms of  RMSE of the ensemble-mean forecast of near-surface temperature,
the effects of AMPT  were mixed.
For near-surface wind speed forecasts, the effects of the atmospheric AMPT 
were similar to but weaker than the effects on near-surface temperature forecasts.
Soil AMPT perturbations imposed  in addition to atmospheric AMPT perturbations
led to nearly uniform further improvements in the ensemble performance
for near-surface temperature and smaller improvements in near-surface wind speed.
The impact of AMPT on precipitation forecast scores was mixed.

In the free atmosphere, AMPT  consistently generated bigger spread than SPPT, without degrading  RMSE. The positive 
effect of AMPT, though, decreased away from the surface.
We recall that our focus in this study was on the near-surface fields, so we did nothing to increase
the spread induced by AMPT in the free atmosphere.


%
%

\section{Conclusions}

A new technique called Additive Model-uncertainty perturbations scaled by Physical Tendency (AMPT)
has been proposed. 
AMPT addresses some issues of the wide-spread SPPT scheme.
First, in contrast to SPPT, AMPT can generate perturbations with significant magnitude even at grid points
where the model physical tendency happens to be spuriously small or is small due to cancellation
of contributions from different physical parametrizations.
Second, AMPT improves on the SPPT's unphysical perfect correlations of perturbations between different model variables
and in the vertical. Third, due to the non-local link from the model fields to the scaling physical tendency, which
determines the magnitude of AMPT perturbations, the SPPT's ban on sign reversal of the physical tendency, Eq.(\ref{sign_rev}), is removed in AMPT. This enables AMPT to generate significantly greater perturbations than SPPT  without causing instabilities.

Taking the unperturbed model run to introduce state-dependence in model perturbations
for ensemble members is discussed. This approach breaks the positive feedback loop from
the model state to perturbations and then back to the model state at a later lead time, 
therefore, it can be used to 
maintain stability of ensemble forecasts perturbed in a state-dependent manner.
We found in this research that just relaxing the pointwise dependence of the model
perturbation magnitude on the model state (by using a spatial moving average of the
magnitude of physical tendency and by updating it less frequently than each model time step) 
helped much in preventing instabilities.

AMPT employs the Stochastic Pattern Generator \citep[SPG,][]{TsyGaySPG} to generate 
four-dimensional random fields with tunable spatio-temporal correlations
(but can be used with any pattern generator). The  random fields 
generated by SPG to perturb  different model fields are mutually independent. They
are scaled by the area-averaged modulus of physical tendency and added to the model fields
at every model time step. 
AMPT perturbations of three-dimensional atmospheric model fields of temperature, pressure
(computed from temperature perturbations via hydrostatics), wind, humidity,
and three-dimensional  soil fields (temperature and moisture)  were imposed and their effects
on convection-permitting ensemble forecasts (based on the COSMO forecast model) were examined.

Practically, AMPT performed better than SPPT in our ensemble prediction 
experiments but at the expense of 
being computationally more  expensive than SPPT.
The bigger cost of AMPT was mostly  because the AMPT's random patterns are three-dimensional (in the atmosphere),
as opposed to two-dimensional random patterns in SPPT.
In  two dimensions (in the horizontal, soil perturbations), the cost of running AMPT was negligibly small. 
The overhead of running AMPT with the perturbed three-dimensional $T,u,v$ fields 
(as well as $p$, $T_{\rm SO}$, and $W_{\rm SO}$)
was about 10\%  of the time of running the COSMO model itself. 
There were three main computer-intensive parts of AMPT.
First, the spectral SPG solver, which was optimized and parallelized, was responsible for 
about 25\% of the total burden of AMPT. 
Second, the three-dimensional fast Fourier transform (FFT) 
took another 25\% of computer time. FFT was only partly optimized and done with an old
legacy code. The third computer intensive part was the 
computation of the scaling physical tendencies ${\cal P}_i$ and spatial interpolations. 
This part was not parallelized at the time of writing this paper.
As a result, AMPT was about 10 times more expensive than 
SPPT, but the gap can be reduced by a factor of three after the AMPT code is fully optimized in preparation
for an operational application.

Another difference between AMPT and SPPT is that AMPT has many more parameters.
On the one hand, this can be viewed as a disadvantage because AMPT requires more
effort to tune them. On the other hand,
all the AMPT parameters have clear physical meaning and are set to meaningful
default values. This offers the AMPT user a choice: either to set the  parameters to their
default values indicated in section \ref{sec_expm}
or to tune some of the parameters taking into
account specifics of a particular application.

It is also worth noting that AMPT suffers from a lack of physical consistency even to a greater extent than SPPT
since AMPT perturbations are not balanced. We argue that this might not be a problem because of a
small magnitude of model perturbations introduced every model time step so that the model manages to balance
them itself. Experimentally, we found no indications of problems related to the lack of balance in AMPT 
perturbations.

The main findings of the study are the following.

\begin{itemize}

\item
Humidity perturbations generated significant 
ensemble spread in precipitation forecasts
but led to a high bias-to-spread ratio.
For this reason, AMPT was systematically tested without humidity perturbations.

\item
Withholding temperature perturbations led to a substantial reduction in the
bias and in the bias-to-spread ratio, but at the expense of a substantial reduction in spread as well.
For this reason, temperature (and pressure) were included in the list of perturbed model fields 
in the final ensemble prediction experiments.

\item
In ensemble prediction experiments, it was found that AMPT 
was much  more effective in generating spread in the ensemble than SPPT,
thus considerably improving {reliability} of the ensemble.
This was the case for the upper-air fields as well as for the near-surface fields.

\item
Most probabilistic ensemble verification scores for near-surface
temperature and wind speed
were improved due to atmospheric AMPT perturbations as compared with SPPT
(at the statistical significance level of 0.05).

\item
Impacts of AMPT on the root-mean-square errors  of the ensemble-mean 
temperature and wind speed forecasts  were  mixed (compared with SPPT).
This held for  free-atmosphere as well as near-surface variables.

\item
AMPT perturbations of soil moisture and soil temperature significantly
improved  deterministic and probabilistic verification scores
for  near-surface temperature and had little effect on near-surface wind speed.

\item
Probabilistic verification of  ensemble forecasts in terms of  precipitation gave mixed results, perhaps, due to an
insufficient number of precipitating events during the time period examined and scarcity of the observation network.

\end{itemize}

The technique can be further developed in the following directions.
First, it looks reasonable to introduce state-dependence not only in the magnitude
of perturbation fields but also in their spatial and temporal scales.
This will require a non-stationary (in space and time) stochastic random field generator
(\eg based on the technique proposed  in \citet{Tsyrulnikov2019})
and a method to specify the non-stationarity patterns. 
Second, in its current formulation, AMPT takes the net physical tendency as input but it 
can be used at the process level as well, \ie for the output of each parametrization scheme separately.
Third,  switching from perturbations of tendencies to perturbations of fluxes looks
as a promising way to achieve physical consistency.
Finally, we note that AMPT can be used in data assimilation  as  well as in ensemble prediction systems:
in the atmosphere, in the soil, and also in the ocean.

\section*{Acknowledgments}

We are grateful to T.Paccagnella and A.Montani, who provided us with initial and boundary 
conditions for our forecast ensemble.
We would like to thank members of the COSMO Working Group on Predictability and
Ensemble Methods for many fruitful discussions.
V.Volkova assisted with the ensemble forecast  verification in the free atmosphere.
A.Kirsanov and A.Bundel kindly helped us to use the VERSUS verification tool.
We also thank D.Blinov for  assistance in working with the COSMO model.
Valuable comments of two anonymous reviewers led to significant improvements of the manuscript.
The study has been conducted
under  Tasks  1.1.1 and  1.1.4 of the Scientific Research and Technology 
Plan of the Russian Federal Service for
Hydrometeorology and Environmental Monitoring.

\section*{Data availability statement}

The dataset on which this paper is based is too large to
be retained or publicly archived with available resources.
Documentation and methods used to support this study
are available from the first author.
The Fortran source code of SPG is available from \url{https://github.com/gayfulin/SPG}. 
Reproducible R code that realizes the methodology of statistical significance testing
(presented in Appendix B), along with raw data,
can be found at \url{https://github.com/cyrulnic/AMPT}.

\section*{Disclosure statement}

No potential conflict of interest was reported by
the authors.

\section*{Authors' contributions}

All authors contributed to the study conception and design. 
Michael Tsyrulnikov wrote the manuscript.
Elena Astakhova performed ensemble prediction experiments and verification.
Dmitry Gayfulin wrote the program code and conducted preliminary sensitivity tests.
All authors commented on previous versions of the manuscript. 
All authors read and approved the final manuscript.


\section*{Appendices}
\addcontentsline{toc}{section}{\numberline{}Appendices}%

\appendix

\section{Additive and multiplicative perturbations}
\label{App_add_mult}

This Appendix contains  remarks on the suitability of the term `additive' in the acronym AMPT.

In stochastic systems theory, the term `additive noise' denotes a perturbation that does not depend 
on the system state. If the perturbation is a function of the system state, it is called multiplicative, \eg \citet[][sec. 6.5]{Fuchs}.
We recall that
in ensemble prediction, an ensemble member can be viewed as a numerical solution to the stochastic dynamic
equation of the unknown, random {\em truth} (system state), \eg \citet[][Eq.(1.3)]{TsyGaySPG}:
${\d {\bf x}}/{\d t}=  {\bf F}({\bf x}) - \boldsymbol\xi$
(where ${\bf x}$ is the truth, ${\bf F}$ is the forecast-model operator, and  $\boldsymbol\xi$ is the random
model perturbation).
So, the model perturbation is multiplicative if it is 
a function of ${\bf x}$ (the state of the ensemble member on which the model perturbation is imposed) 
and additive if it does not depend on ${\bf x}$.

With SPPT, the model perturbation at some point in space and time, $({\bf s}, t)$, 
is proportional to the physical tendency ${\bf P}({\bf x}({\bf s}, t))$,
where ${\bf x}$ is the state of the ensemble member, Eq.(\ref{d_SPPT}). 
Therefore,  the SPPT perturbation in each model variable $i$  
is a function of ${\bf x}$ and thus is multiplicative:
\begin {equation}
\label{SPPT_pert}
\Delta P_i^{\rm SPPT}({\bf s}, t) =  \kappa \, P_i({\bf x}({\bf s}, t)) \, \xi({\bf s}, t).
\end {equation}
With AMPT, if the scaling tendency is taken from the control (unperturbed) run, so that
${\cal P}_i({\bf s}, t) = {\cal P}_i({\bf x}^{\rm control}({\bf s}, t))$, 
the perturbation ceases to depend on ${\bf x}$:
\begin {equation}
\label{AMPT_pert}
\Delta P_i^{\rm AMPT}({\bf s}, t) =  
  \kappa_i \, {\cal P}_i({\bf x}^{\rm control}({\bf s}, t)) \, \xi_i({\bf s}, t),
\end {equation}
it only depends on the {\em known} initial conditions of the control run. Thus, the AMPT perturbation
is indeed additive in the setting in which the scaling physical tendency is taken from the control model run.

\section{Methodology of testing statistical significance}
\label{App_signif}

Here we describe how we calculated statistical significance of improvements due to AMPT
in ensemble prediction experiments presented in section
\ref{sec_expm_eps}.

Let us test the hypothesis that a verification score $\psi$ 
is greater for scheme 1 than for scheme 2.
More precisely, 
we are going to compare the scores $\psi_1$ and $\psi_2$ 
that could be obtained from a very large sample of forecasts,
whilst having their unbiased estimates $\widehat\psi_1$ and $\widehat\psi_2$ obtained from smaller samples.
That is, the hypothesis to be tested is $H: \psi_1 \equiv \Ex\widehat\psi_1 > \Ex\widehat\psi_2 \equiv \psi_2$.
The default  `null' hypothesis states that there is no difference between the two schemes: 
$H_0: \psi_1=\psi_2$.

Since atmospheric states are correlated in time and forecast errors depend on the atmospheric flow,
forecast errors are also temporally correlated (dependent). Accounting for these time correlations would
complicate statistical hypothesis testing, so for simplicity, we employ the following approach.
We divide the whole two-month period into $n=8$ weeks and compute the score $\psi$ for each of the two schemes 
for each week separately, getting the respective sequences
$\widehat\psi_{1j}$ and $\widehat\psi_{2j}$ (for $j=1,2,\dots,n$)
and their differences $\delta_j^{\rm obs} = \widehat\psi_{1j} - \widehat\psi_{2j}$.
We treat $\delta_j^{\rm obs}$ as (observed) realizations of the respective random variables $\delta_j$. 
Assuming that  temporal dependencies in $\delta_j$
decay on a time scale shorter than one week, we
regard $\delta_j$ as a sample of statistically independent and identically distributed random numbers.

Informally, if $\delta_j^{\rm obs}$ are mostly positive, we tend to believe that $\Ex \delta_j>0$, 
meaning that the hypothesis $H$ we are testing is true.
Formally, we reject $H_0$ (and thus accept $H$) if 
$p=P(\bar\delta \ge \bar\delta^{\rm obs} \mid H_0)$, where the overbar denotes sample averaging, is sufficiently small.
This means that if, under $H_0$, the probability of the observed (and more extreme) 
deviation of $\bar\delta$ away from zero is small, then the data we have are, likely, not consistent with $H_0$.
Specifically, we select a {\em significance level}, $\alpha$, and reject $H_0$
if the above probability $p$ (known as the $p$-value) is less than  $\alpha$.

To compute $p$ we need to know the probability distribution of $\delta_j$ under $H_0$.
This can be easily done using an  approach to statistical inference known as bootstrap.
In its simplest flavor, bootstrap
postulates that the data distribution (\ie the distribution of $\delta_j$) equals the empirical distribution 
(which is concentrated at the observed data, in our case, $\delta_j^{\rm obs}$).
However, this setup cannot be directly applied here because $H_0$ assumes that $\Ex \delta_j=0$
whereas the empirical distribution is, likely, biased ($\bar\delta^{\rm obs} \ne 0$). 
Following \citet[][ch.16]{Efron}, to define the bootstrap distribution, we shift the
empirical distribution so that it has zero mean.
Then, we proceed as usual, sampling (with replacement) from this discrete distribution
(concentrated with equal probabilities at the points $\Delta_j^{\rm obs} = \delta_j^{\rm obs} - \bar\delta^{\rm obs}$),
compute the mean in each bootstrap sample, $\bar\delta^{*}$, and calculate the fraction of bootstrap samples in which 
$\bar\delta^{*} > \bar\delta^{\rm obs}$. This fraction is the estimate of the $p$-value.


\bibliography{mybibfile}


%
%

\end{document}